\documentclass[aps,pra,preprint, amsmath, nolongbibliography]{revtex4-2}

\usepackage{pbox}
\usepackage{array}
\usepackage[T1]{fontenc}
\usepackage{graphicx,color}
\usepackage{bm}        % \boldsymbol (bold math)
\usepackage{amssymb}   % \mathbb     (blackboard bold)
\usepackage{amsmath} 
\usepackage[capitalize]{cleveref}
\usepackage{amsmath,amssymb,amsfonts, bm} %Mathematical Symbol Packages
\usepackage{graphicx} %Figures Packages

\newcommand{\beq}{\begin{equation}}
\newcommand{\eeq}{\end{equation}}
\newcommand{\bra}[1]{\langle #1 |}
\newcommand{\ket}[1]{| #1 \rangle}
\newcommand{\beqs}{\begin{subequations}}
\newcommand{\eeqs}{\end{subequations}}
\newcommand{\nbar}{\bar{n}}

\newcommand{\half}{\frac{1}{2}}

\begin{document}
\date{\today}
\flushbottom

\title{Population Oscillations and Ubiquitous Coherences in multilevel quantum
systems driven by incoherent radiation}
\author{Amro Dodin}
\email{adodin@lbl.gov}
\affiliation{Chemical Sciences Division, Lawrence Berkeley National Laboratory, Berkeley, CA 94720, USA}
\author{Timur V. Tscherbul} 
\affiliation{Department of Physics, University of Nevada, Reno, NV, 89557, USA}
\author{Paul Brumer}
\affiliation{Chemical Physics Theory Group, Department of Chemistry, and Center for Quantum Information and Quantum Control, University of Toronto, Toronto, Ontario, M5S 3H6, Canada}

\begin{abstract}
   We consider incoherent excitation of multilevel quantum systems, e.g. molecules with multiple vibronic states.
   We show that (1) the geometric constraints of the matter-field coupling operator guarantee that noise-induced coherences will be generated in all systems with four or more energy eigenstates and (2) noise-induced coherences can lead to population oscillations due to quantum interference via coherence transfer between pairs of states in the ground and excited manifolds. Our findings facilitate the experimental detection of noise-induced coherent dynamics in complex quantum systems.
    
\end{abstract}

\maketitle

In quantum devices, noise is typically a nuisance that disrupts coherent dynamics and must be mitigated to maintain performance in high-precision measurement \cite{beloy_high-accuracy_2006,ovsiannikov_rydberg_2011,safronova_blackbody_2012,lisdat_blackbody_2021,ajoy_atomic-scale_2015,stanwix_coherence_2010,bar-gill_suppression_2012,degen_quantum_2017,dolde_electric-field_2011,doherty_nitrogen-vacancy_2013,neumann_high-precision_2013, zhang_quantum-assisted_2022} and quantum computing \cite{nielsen_quantum_2010,monroe_quantum_2002,preskill_quantum_2018,resch_benchmarking_2021, dawson_noise_2006, bharti_noisy_2022}.
However, this same noise, in the form of sunlight or heat, can excite  highly nontrivial superposition states \cite{dodin_noise-induced_2022,dodin_secular_2018,koyu_steady-state_2021,koyu_long-lived_2022}, displaying noise-induced coherences that modify fluorescence emission \cite{dodin_secular_2018,koyu_steady-state_2021,koyu_long-lived_2022}, enhance the power output of quantum heat engines \cite{scully_quantum_2011,gelbwaser-klimovsky_power_2015}, and alter the non-adiabatic relaxation  and quantum yield of retinal photoisomerization \cite{tscherbul_excitation_2014,tscherbul_quantum_2015, dodin_light-induced_2019}.
While noise-induced coherences have been studied in many model systems, it remains unclear if they can arise in any quantum system under the right conditions, or if they are only possible in a few special cases.
In this Letter, we show that noise-induced coherences are generated by incoherent excitation of all but the simplest systems, a consequence of geometric constraints of the light-matter coupling Hamiltonian.
This surprising result demonstrates that noise-induced coherences are ubiquitous and not a consequence of special system properties.
Furthermore, we show that, similarly to coherent laser pulses, noise excitation can produce coherent population oscillations due to interference.

Oscillatory dynamics are important because they (1) demonstrates that oscillatory interference requires neither a well-defined phase of  the exciting field nor impulsive excitation,  and (2) predicts that noise-induced coherence can lead to population dynamics that differ \textit{qualitatively} from classical rate models - facilitating laboratory demonstration of noise-induced coherences.
The coherent population oscillations are a new dynamical regime in noise-driven dynamics and arise due to a novel coherence transfer mechanism where the phase of the noise generated excited state superposition depends on the phase of the pre-excitation ground state superposition.
Such a mechanism could  not occur in previously studied three level models since it requires a minimum of four states, two in each manifold, and could not be identified in larger models, \cite{dodin_light-induced_2019, tscherbul_quantum_2015} due to their complexity.

\begin{figure}[htbp]
	\centering
	\includegraphics[width=0.9\columnwidth]{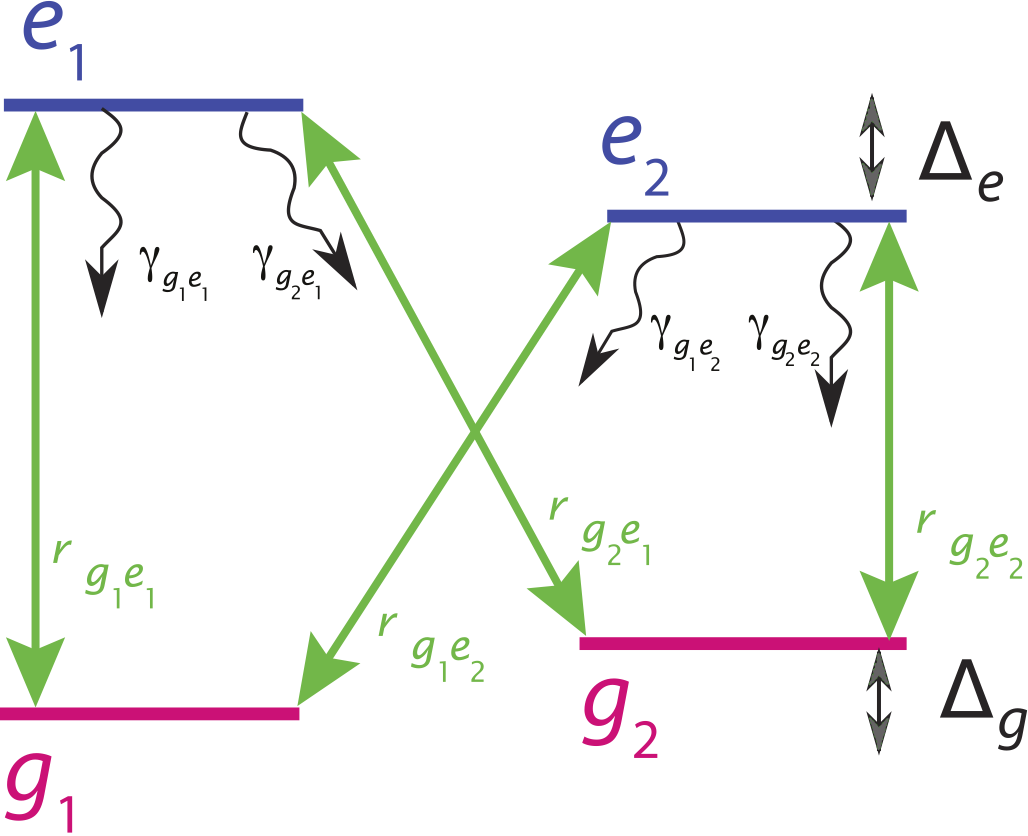}
	\renewcommand{\figurename}{Fig.}
	\caption{Schematic of the four-level model system under consideration. Ground eigenstates $g_1$ and $g_2$ are coupled to excited  eigenstates $e_1$ and $e_2$ by incoherent light. Ground and excited state pairs are separated by energies $\Delta_g$ and $\Delta_e$ respectively. Ground state $g_i$ is driven into excited state $e_j$ with incoherent pumping rate $r_{g_ie_j}$. The excited state decays to ground state through spontaneous and stimulated emission at rates $\gamma_{g_ie_j}$  and $r_{g_ie_j}$ respectively.}
\label{fig:4LS}
\end{figure}

We demonstrate these features in a four level system (4LS) with two ground ${g_i}$ and two excited ${e_i}$ states, shown in Fig. \ref{fig:4LS}.
The ground (excited) states are separated by the energy gap $\Delta_g$ ($\Delta_e$).
The two manifolds are separated by the energy gap $\Delta_0 \gg \Delta_g, \Delta_e$.
The 4LS is excited by a thermal bath coupled to its dipole moment operator which drives ground state $g_i$ to excited state $e_j$ with rate $r_{g_ie_j}$.
The excited state ${e_j}$ decays to ground state ${g_i}$ through spontaneous (stimulated) emission with rate $\gamma_{g_ie_j}$ ($r_{g_ie_j}$).
The 4LS lacks any cycles of states (e.g. of the form $A\to B\to C\to A$) which could produce oscillatory population dynamics classically.
Thus, any population oscillations arise due to quantum interference.

The partial secular Bloch-Redfield (PSBR) equation \cite{tscherbul_partial_2015} for this model is given (in $\hbar=1$ units) by
\begin{widetext}
\begin{subequations}
\label{eq:QME}
\begin{equation}
\label{eq:epop}
\dot{\rho}_{e_ie_i}= \sum_{j=1}^2 \left[r_{g_je_i}\rho_{g_jg_j} -(1+\nbar)\left(\gamma_{g_je_i}\rho_{e_ie_i}+p_{g_j}\sqrt{\gamma_{g_je_1}\gamma_{g_je_2}}\rho_{e_1e_2}^R\right)\right] +2p_{e_i}\sqrt{r_{g_1e_i}r_{g_2e_i}}\rho_{g_1g_2}^R
\end{equation}
\begin{equation}
\label{eq:gpop}
\dot{\rho}_{g_ig_i}=\sum_{j=1}^2 \left[(1+\nbar) \gamma_{g_ie_j}\rho_{e_je_j} -p_{e_j}\sqrt{r_{e_jg_1}r_{e_jg_2}}\rho_{g_1g_2}^R -r_{g_ie_j}\rho_{g_ig_i}\right]
+2(1+\nbar)p_{g_i}\sqrt{\gamma_{g_ie_1}\gamma_{g_ie_2}}\rho_{e_1e_2}^R
\end{equation}
\begin{equation}
\label{eq:ecoh}
\begin{split}
\dot{\rho}_{e_1e_2} = &
\sum_{j=1}^2 \left[ p_{g_j}\sqrt{r_{g_je_1}r_{g_je_2}}\rho_{g_jg_j} -\half(1+\nbar)p_{g_1}\sqrt{\gamma_{g_je_1}\gamma_{g_je_2}}(\rho_{e_1e_1}+\rho_{e_2e_2})\right]\\
&-\half(1+\nbar)\sum_{i,j=1}^2\gamma_{g_je_i}\rho_{e_1e_2}-i\Delta_e\rho_{e_1e_2}+(p_{||}\sqrt{r_{g_1e_1}r_{g_2e_2}}+p_{\times}\sqrt{r_{g_1e_2}r_{g_2e_1}})\rho_{g_1g_2}
\end{split}
\end{equation}
\begin{equation}
\label{eq:gcoh}
\begin{split}
\dot{\rho}_{g_1g_2}=& \sum_{j=1}^2\left[(1+\nbar)p_{e_j}\sqrt{\gamma_{g_1e_j}\gamma_{g_2e_j}}\rho_{e_je_j} -\half p_{e_j}\sqrt{r_{g_1e_j}r_{g_2e_j}}(\rho_{g_1g_1}+\rho_{g_2g_2})\right]\\
&+ (1+\nbar)\left(p_{||}\sqrt{\gamma_{g_1e_1}\gamma_{g_2e_2}}+p_{\times}\sqrt{\gamma_{g_2e_1}\gamma_{g_1e_2}}\right)\rho_{e_1e_2} 
-\half\sum_{i,j=1}^2r_{e_ig_j}\rho_{g_1g_2}-i\Delta_g\rho_{g_1g_2},
\end{split}
\end{equation}
\end{subequations}
\end{widetext}
where $\rho_{ij}^R \equiv (\rho_{ij}+\rho_{ji})/2$ is the real part of coherence, and the decay and excitation rates are $\gamma_{g_ie_j}=\omega_{g_ie_j}^3|\mu_{g_ie_j}|^2/(3\pi\epsilon_0c^3)$ and $r_{g_ie_j}=\gamma_{g_ie_j}\nbar(\omega_{g_ie_j})$.
The rates depend on the mean bath occupation $\nbar(\omega)=[\exp(\hbar\omega/k_BT)-1]^{-1}$ and the transition dipole moments $\bm{\mu}_{ij}$.
The generation of noise-induced coherences, and their coupling to populations via interference are controlled by the alignment parameters $p_\alpha=\bm{\mu}_{i\alpha}\cdot\bm{\mu}_{\alpha l}/(|\mu_{i\alpha}||\mu_{\alpha l}|)$ as discussed in more detail in appendix \ref{sec:PSBR}. We note here the new coherence transfer terms coupling $\rho_{g_1g_2}$ and $\rho_{e_1e_2}$ which are absent in previously studied 3LS's and will be crucial to the phenomenology of the 4LS.

\textit{Ubiquity of Coherences:} The generation of coherence and subsequent interference is controlled by six alingnment parameters $\lbrace p_\alpha \rbrace$, which depend on the geometry of $\bm{\mu}_{ij}$.
Noise-induced coherences will be generated whenever any $\lbrace p_\alpha \rbrace$ parameters are non-zero and are maximized as $p_\alpha\to1$.
Thus, the only systems that lack noise-induced coherence are those where all transition dipole moments are mutually orthogonal.
In three dimensional (3D) physical space, this is only possible when there are three or fewer transitions.
In other words, \textit{any quantum system with 4 or more dipole-allowed transitions  will generate noise-induced coherence, because 4 or more 3D vectors cannot be mutually orthogonal, indicating that these coherences are ubiquitous in realistic systems}.
%While partial secular equations are completely positive when properly derived from an interaction Hamiltonian \cite{dodin_secular_2018, trushechkin_unified_2021, farina_open-quantum-system_2019}, failure to satisfy alignment constraints yields a system with no corresponding dipole-interaction Hamiltonian and the QME can yield unphysical dynamics.

These insights are not restricted to photo-induced processes.
System-phonon interactions are also proportional to a system position operator \cite{svidzinsky_enhancing_2011}.
Therefore, they are subject to the same constraints as the dipole operator and will produce ubiquitous noise-induced coherences.
These arguments rely only on the structure of the system-bath interaction Hamiltonian, independent of the bath Hamiltonian and temperature, and thus also apply to non-equilibrium systems interacting with multiple baths.
In equilibrium, noise-induced coherences are only important in the transient response of the system, since it will eventually reach an incoherent thermal steady state.
However, when the system is coupled to multiple baths, the non-equilibrium steady state (NESS) may be coherent, with non-zero coherences which can alter energy transfer rates and heat currents through the system \cite{koyu_steady-state_2021,jung_energy_2020,joubert-doriol_quantum_2023,ivander_quantum_2022}.
Coherent NESS have been seen when the system (1) has some non-zero $p$ alignment parameters, and (2) is coupled to several baths at different temperatures \cite{dodin_light-induced_2019}.
The arguments here address the first condition, indicating that systems capable of supporting coherent NESS when coupled to suitable baths are ubiquitous.
The second condition is a statement on bath fluctuations and can not be addressed through the system-bath coupling alone.
Therefore, whereas systems that \textit{can} support coherent NESS are ubiquitous, combinations of baths that \textit{produce} coherent NESS is the subject of a separate study.

The ubiquity of noise-induced coherences seems inconsistent with the success of secular rate law equations.
This apparent inconsistency is resolved by noting that, although noise-induced coherences arise as a necessary consequence of incoherent excitation, they significantly affect observables only in systems with pairs of nearly-degenerate states and nearly-parallel transition dipole moments.
In such systems, the same radiation field modes can drive two different transitions, allowing them to interfere.
Similar conditions were noted in the coherent control of internal conversion in pyrazine \cite{grinev_coherent_2015,christopher_overlapping_2005,christopher_quantum_2006, christopher_efficient_2006}, where interference in radiationless relaxation was only possible between states with overlapping resonances.

The dynamics of the 4LS has two regimes in each manifold \cite{tscherbul_long-lived_2014, dodin_quantum_2016}, determined by their energy spacing $\Delta$ and inverse lifetime $\Gamma = \tau^{-1}$.
Excited state lifetimes are determined by $\Gamma=\gamma (1+\nbar)\approx\gamma$ for most incoherent excitations, where $\nbar(\Delta_0)\ll 1$ while ground state lifetimes are determined by $\Gamma=r = \gamma \nbar\ll \gamma$.
In the underdamped ($\Delta > \Gamma$) regime, the states display weak oscillatory coherences that do not couple to populations.
In the overdamped ($\Delta <\Gamma$) regime, coherences can be large in magnitude and couple strongly to populations.
They are no longer oscillatory and rise to a coherent quasi-stationary superposition state before decaying into an incoherent thermal state.
Figure \ref{fig:CohExample} shows representative examples of coherence dynamics in these regimes.
As confirmed in appendix \ref{sec:Numerical}, this behavior is consistent with expectations from three level models.
The light is turned on at t=0 and stays on thereafter (Differences from incoherent pulses are discussed in appendix \ref{sec:pulse}).
%The dynamics were obtained by direct numerical integration of the BR equations.

\begin{figure}[htbp]
	\centering
	\includegraphics[width=\columnwidth]{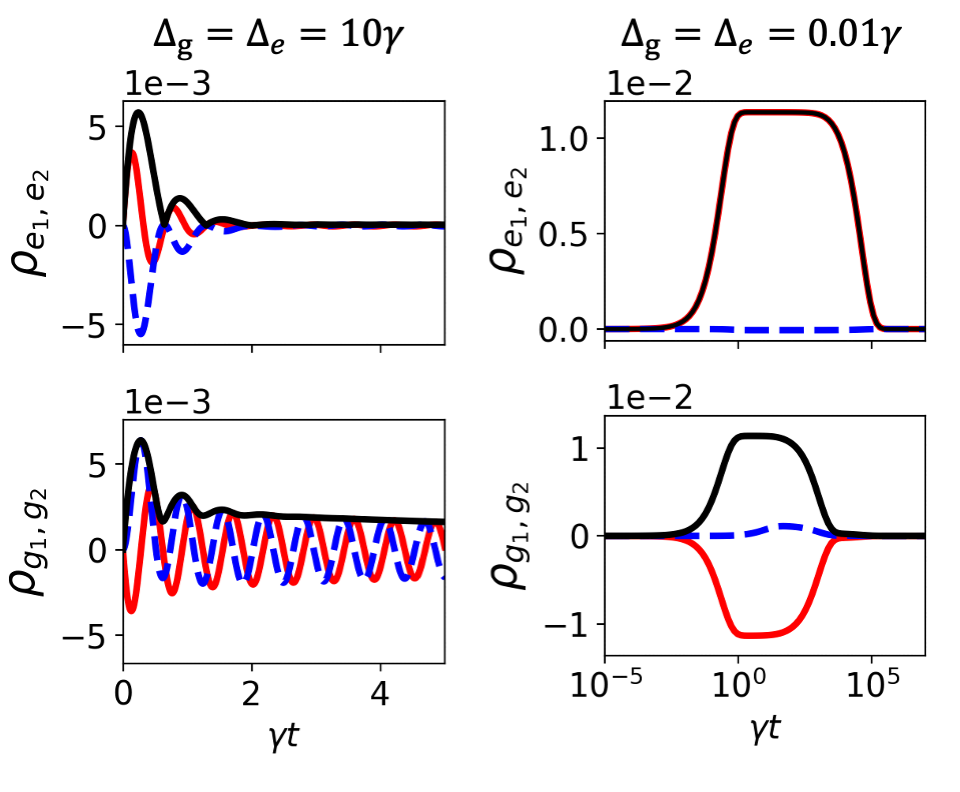}
	\renewcommand{\figurename}{Fig.}
	\caption{Dynamics of the real (red solid), imaginary (blue dashed) and magnitude (black solid) of coherences in each manifold. Time is reported in terms of the mean spontaneous emission rate $\gamma$. All transitions have the same rates $\gamma_{g_ie_j}=\gamma$ and $r_{g_ie_j}=\nbar\gamma$ with $\nbar=0.05$. All transition dipole moments are taken to be parallel with alignment parameters $p_i=1$.}
\label{fig:CohExample}
\end{figure}

\begin{figure}[htbp]
	\centering
	\includegraphics[width=\columnwidth]{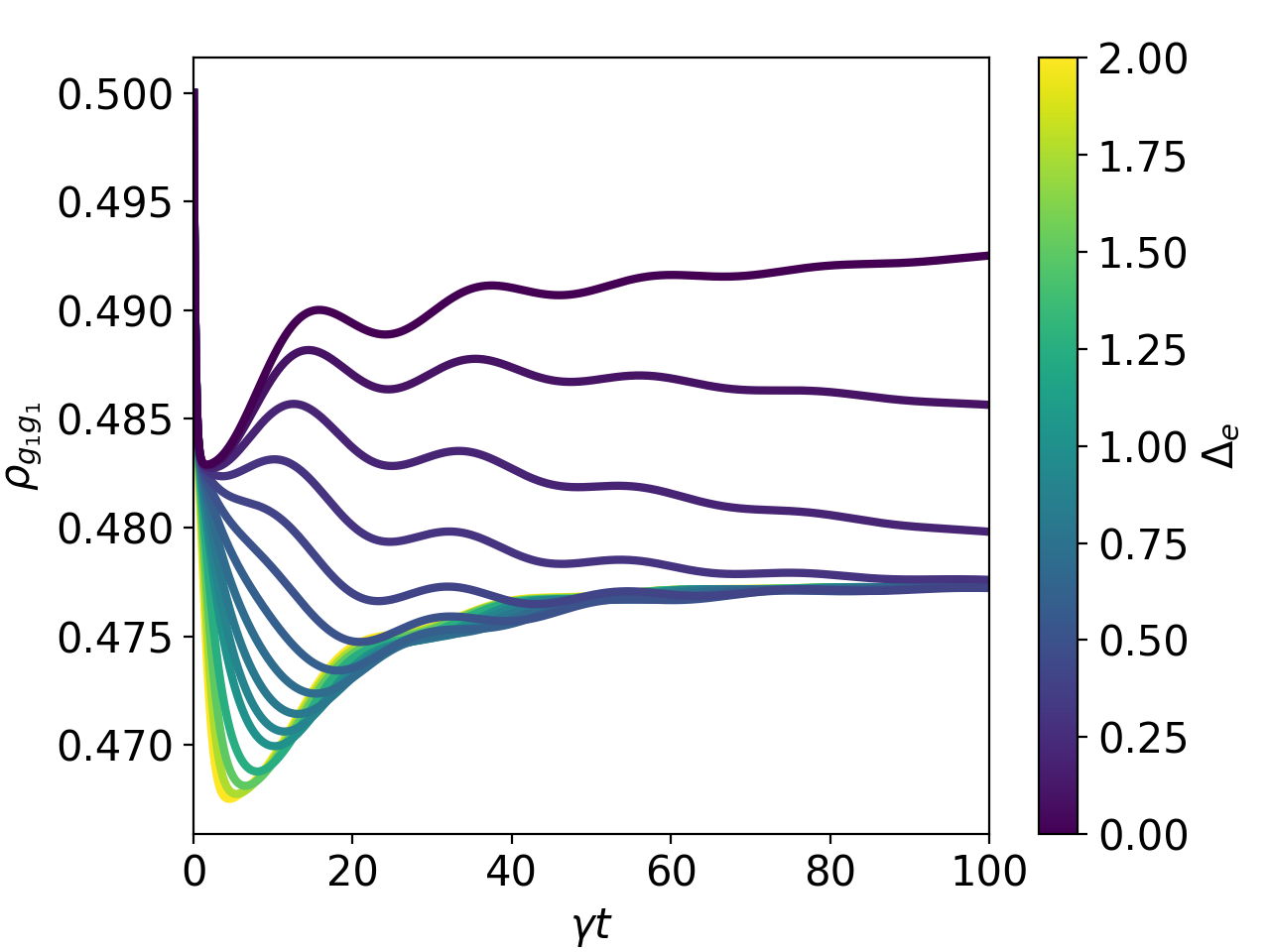}
	\renewcommand{\figurename}{Fig.}
	\caption{Coherent ground state population oscillations driven by incoherent light in  the mixed damping regime. Coherent oscillations in ground state  population depend on excited state splitting $\Delta_e$.  The colored traces show a range of  excited state splittings $\Delta_e=0$ (the limit of overdamped excited state coherences) to $\Delta_e=2\gamma$, entering the underdamped regime. Further increasing $\Delta_e$ does not appreciably change the plots. In all cases, $\Delta_g=0.3\gamma=6r$ ($\nbar=0.05$) is in the underdamped regime. Ground state couplings are $\gamma_{g_1e_j}=1.5\gamma$ and $\gamma_{g_2e_j}=0.5\gamma$.}
\label{fig:OUDInterference}
\end{figure}

\textit{Coherent Population Oscillations:} 
This separation of dynamics into two regimes in three level models suggests that noise excitation cannot produce coherent population oscillations.
However, the four level model supports one additional scenario - the mixed damping case where one pair of states is underdamped and the other overdamped.
The dynamics of  $\rho_{g_1g_1}$ is shown in Fig. \ref{fig:OUDInterference}, when the ground state is in the underdamped regime, and the excited state splitting $\Delta_e$ is swept between the overdamped and underdamped regimes.
The system is initialized in the ground state mixture $\rho = \ket{g_1}\bra{g_1} + \ket{g_2}\bra{g_2}$ so that, any coherences are generated over the course of noise-driven dynamics.
The mean emission (excitation) rate is  $\gamma \equiv \sum_{i,j} \gamma_{g_ie_j}/4$  ($r = \gamma\nbar$).
The two ground states have different coupling strengths, with $\gamma_{g_1e_j} = 1.5\gamma$ and $\gamma_{g_2e_j}=0.5\gamma$ and fixed energy splitting $\Delta_g=6r$, while the excited state splitting is swept from degeneracy ($\Delta_e=0$) to the onset of the underdamped regime ($\Delta_e = 2\gamma$).
%The populations remain positive and normalized at all times, reproduce the expected thermal steady state, and the coherences satisfy the physical bound $|\rho_{ij}|^2 \leq \rho_{ii}\rho_{jj}$.

The dynamics in Fig. \ref{fig:OUDInterference} are striking.
When the ground state is in the underdamped regime and the excited state is in the overdamped regime, ground state populations show significant coherent oscillations.
As the excited state enters the underdamped regime, these oscillations vanish.
These oscillatory dynamics are attributed to interference since they vanish when coherences are suppressed.
Moreover, as shown in section 2 of the SM , in contrast to  pulsed incoherent light, they cannot be explained by the same mechanism as coherent excitation.
The influence of excited state energy splitting on ground state population dynamics suggests that these oscillations occur due to interference between the $e_j \to g_1$ emission pathways.
However, the frequency of the oscillations does not follow $\Delta_e$ - remaining fixed at the ground state frequency, indicating the role of coherence oscillations in the ground state.
The lifetime of these coherent oscillations is determined by the radiative decay rate, $\gamma$ of the system.
In a molecular system excited by sunlight in the absence of non-radiative relaxation pathways, $\gamma^{-1} \sim 1$ ns, indicating that coherent population oscillations survive for tens of nanoseconds.

The observed coherent population oscillations arise due to a cooperative effect between ground and excited state coherences.
The effect is most pronounced when the excited state coherences couple strongly to population dynamics through interference effects in excitation and emission pathways but, by themselves, display non-oscillatory sigmoid dynamics.
The oscillatory ground state coherences do not couple strongly to the populations but can transfer to the excited state through the coherence transfer terms, where they lead to interference in the emission process from the excited state to the ground states, producing the observed coherent population oscillations.
As $\Delta_e$ increases, the coupling between the excited state coherence and populations decreases, leading to the disappearance of the coherent population oscillations.
%Interestingly, these oscillations are only observed when the ground state couplings are asymmetric.
Moreover,the excited state populations undergo much smaller oscillations than the ground states \ref{sec:Numerical}.
The interference therefore does not change the \textit{total} emission rate from the excited state as much as it changes which ground state receives the relaxing population.

These population oscillations present a new opportunity for the experimental observation of noise-induced coherences.
While the previously predicted coherence effects have suggested deviations from classical bounds, their dynamics are essentially given by rate law dynamics.
In contrast, the population oscillations predicted in this letter display qualitatively different population dynamics that are easier to distinguish than changes in transition rates.
In particular, Fig. \ref{fig:OUDInterference} suggests a simple Transient Absorption experiment for the direct observation of these coherent effects.
An ensemble of molecular or atomic systems with parameters in the mixed damping regime can be continuously illuminated by incoherent light suddenly turned on at $t_0=0$.
Then, the absorption of the ensemble is probed at a time interval $\Delta t\sim 1$ ns after the sudden turn on of the incoherent light at a frequency corresponding to a transition out of the ground state $g_1$.
This transient absorption spectrum will then display an oscillation in the intensity due to the oscillations in the ground state bleach as the population of $g_1$ undergoes coherent population oscillations.
 Ultracold polyatomic molecules and Rydberg atoms are promising systems, in which to detect noise-coherent coherent dynamics due to their intrinsically multilevel structure and sensitivity to blackbody radiation (which is particularly extreme for Rydberg atoms) \cite{augenbraun_direct_2023,gallagher_rydberg_1994,tscherbul_coherent_2014}. Recent experiments have already observed incoherent population transfer between the rovibrational levels of ultracold CaOH molecules driven by room-temperature blackbody radiation \cite{vilas_blackbody_2023,vilas_blackbody_2023-1}, and could likely be extended toward coherence detection.

While these oscillatory dynamics are striking and demonstrate a new physical phenomenology, they are unlikely to play an important role in the natural excitation of, e.g., biomolecules by sunlight.
The oscillations arise in the transient regime, shortly after the sudden turn on of incoherent light but disappear in the equilibrium or non-equilibrium steady states, which dominate the properties of biological systems \cite{axelrod_efficient_2018, axelrod_multiple_2019}.
In addition, the observation of these oscillations requires the sudden turn on of the exciting light.
The instant where time translation symmetry is broken by the sudden increase in the light intensity that determines when oscillations start.
If the intensity is turned on slowly, the molecular response will average over oscillations starting at different times, washing out population oscillations.
This loss of transient dynamics with frequencies shorter than the inverse turn on time $\tau_r^{-1}$ is guaranteed by the recently derived adiabatic modulation theorem \cite{dodin_generalized_2021}.
These oscillations will only be seen if the light intensity changes faster than $\sim 1-10$ nanoseconds, while intensities vary on much longer timescales in nature (e.g. the blink of an eye is roughly 0.5 ms $\gg 10$ ns).
Transient responses, however are not entirely irrelevant in the steady state.
They capture the system's response to rapid perturbations, e.g. the excursions from the NESS studied in Ref. \cite{joubert-doriol_quantum_2023}.
Specifically, if a system in the NESS experiences a rapid change in intensity, it may transiently display the same types of population oscillations.

We have demonstrated that in all but the simplest systems, coherences are ubiquitous and are generated as a natural consequence of excitation by incoherent fields.
Therefore, the generation of noise-induced coherence requires no special molecular properties, and is instead required by the geometry of light-matter coupling.
This does not imply that coherent effects must always contribute strongly, as the generated coherences may be small or short-lived.
They contribute most strongly between nearly degenerate energy eigenstate with non-orthogonal transition dipole moments.
In addition, we have shown how coherence transfer between ground and excited states can lead to oscillatory interference in emission rates and consequently coherent population oscillations in the mixed damping regime.
These transient oscillations survive for a period proportional to the mean radiative lifetime and are unlikely to contribute to biomolecular dynamics where the steady state dominates and light intensities vary slowly on molecular timescales.
However, they are surprisingly long lived and open the door to the verification of noise-induced coherence by transient absorption experiments that probe oscillations in the ground state bleach upon incoherent excitation.
Moreover, they call into question the intuition that coherent population oscillations can arise only due to well-defined exciting field phase or specified impulsive excitation time,  revealing a new mechanism by which quantum dynamics can influence population dynamics.

\begin{acknowledgments}
    This work was supported by the U.S. Air Force Office of Scientific Research (AFOSR) under contract No. FA9550-20-1-0354.
\end{acknowledgments}

\appendix
\section{Partial Secular Bloch Redfield Equations}
\label{sec:PSBR}

We consider the dynamics under a partial secular approximation \cite{tscherbul_partial_2015, trushechkin_unified_2021} where we neglect the coherences between ground and excited states that are separated by the large energy gap $\sim \Delta_0$  but retain coherences between the more closely spaced states within the same manifold.
Applying a further Born-Markov approximation yields the following Partial-Secular Bloch-Redfield (PSBR) Master Equations for the system density matrix\cite{tscherbul_partial_2015}, given by Eq. \ref{eq:QME}.
The real part of the coherences is denoted by $\rho_{ij}^R \equiv (\rho_{ij}+\rho_{ji})/2$.
The rates defined above are given by $\gamma_{g_ie_j}=\omega_{g_ie_j}^3|\mu_{g_ie_j}|^2/(3\pi\epsilon_0c^3)$ and $r_{g_ie_j}=\gamma_{g_ie_j}\nbar(\omega_{g_ie_j})$.
The permittivity of free space is $\epsilon_0$ and $\nbar(\omega)=[\exp(\hbar\omega/k_BT)-1]^{-1}$ is the mean occupation number of a bosonic mode with frequency $\omega$.
We have assumed that the occupation number of the radiation field does not vary significantly among the transition frequencies between the ground and excited states, $\nbar(\omega_{g_ie_j}) \approx \nbar$ for all transitions in the system (the Wigner-Weisskopff approximation) \cite{scully_quantum_1997}.
This approximation is well satisfied, e.g. for sunlight driving ground to excited state transitions with frequencies $\omega_0$ in the visible spectrum. 
The $p=\bm{\mu}_{ij}\cdot\bm{\mu}_{kl}/(|\mu_{ij}||\mu_{kl}|)$ coefficients describe the relative orientation of the transition dipole moments, $\bm{\mu}_{ij}=\bra{i}{\bm{\mu}}\ket{j}$, between a given pair of transitions, and play a crucial role in controlling the non-secular effects \cite{tscherbul_partial_2015}.
They quantify the probability that a randomly chosen polarization of the exciting field will simultaneously couple to two different transitions $i\to j$ and $k\to l$.
The alignment parameters labeled by a state, e.g. $p_{e_1}$, define the alignment of the two transition dipole matrices from the labeled state in the excited (ground) manifold to the two states in the ground (excited) manifold.
For example, $p_{e_1}=(\bm{\mu}_{e_1g_1}\cdot\bm{\mu}_{e_1g_2})/(|\mu_{e_1g_1}||\mu_{e_1g_2}|)$.
The parameters $p_{||}=(\bm{\mu}_{e_1g_1}\cdot\bm{\mu}_{e_2g_2})/(|\mu_{e_1g_1}||\mu_{e_2g_2}|)$ and $p_{\times}=(\bm{\mu}_{e_2g_1}\cdot\bm{\mu}_{e_1g_2})/(|\mu_{e_2g_1}||\mu_{e_1g_2}|)$ give the alignment of the transition dipole moments with no shared state.
A detailed derivation of this equation for the general case can be found in Ref. \cite{tscherbul_partial_2015}.

The unitary evolution of the relative phase between energy eigenstates is given by $i\Delta_{g,e}$ terms in the coherence equations of motion, Eqs. (\ref{eq:ecoh}) and (\ref{eq:gcoh}).
The terms coupling populations are the Pauli rate law terms describing excitation, spontaneous emission and stimulated emission between energy eigenstates with no interference.
This incoherent transfer of population between states also causes the decoherence described by the remaining (real) $\rho_{g_1g_2}$ terms in Eq. (\ref{eq:gcoh}) and $\rho_{e_1e_2}$ terms in Eq. (\ref{eq:ecoh}).
Taken together these three types of terms describe the secular evolution of the system which would have been obtained if coherences and populations were decoupled.

However, populations and coherences are coupled, resulting in the generation of noise-induced coherences and associated interference between energy eigenstates.
We provide a brief summary of these terms here to complement the comprehensive physical interpretation presented in Ref. \cite{dodin_noise-induced_2022} and its explanation of how uncorrelated white noise can produce coherences with definite phase.
Simultaneous excitation from a ground state $g_i$ to both excited states generates excited state coherence represented by the terms proportional to $\rho_{g_ig_i}$ in Eq. (\ref{eq:ecoh}).
These coherences can then lead to interference in excitation and decay of the excited states introducing terms proportional to $\rho_{e_1e_2}^R$ in the population equations of motion, Eqs. (\ref{eq:epop}) and (\ref{eq:gpop}).
Moreover, the acceleration in the decay rates due to interference can also accelerate coherence loss in the excited state yielding terms proportional to $\rho_{e_ie_i}$ in Eq. (\ref{eq:ecoh}).
These nonsecular terms involve the simultaneous excitation from or decay to a single ground state ${g_i}$ and are proportional to the alignment parameter $p_{g_i}$.
The physics of these terms has been studied extensively in three level $V$-system models \cite{tscherbul_long-lived_2014,dodin_coherent_2016, dodin_quantum_2016}.
Similar terms arising from simultaneous incoherent excitation to or decay from an excited state ${e_i}$ also appear in the equations of motion and are proportional to the excited state alignment parameters $p_{e_i}$ and have been studied in a three level $\Lambda$-system model \cite{koyu_long-lived_2022}.

However, additional phenomena can appear in larger systems.
For example, consider our four level system.
Specifically, when the system is in an excited state superposition, the state ${e_1}$ can decay to ${g_i}$ and simultaneously ${e_2}$ can decay to $g_{j\neq i}$.
This simultaneous decay between pairs of excited states to different ground and the analogous excitation process from pairs of ground states transfers coherence between the ground and excited states yielding coherence transfer terms proportional to $p_{||}$ and $p_\times$.
We shall see below that \textit{these coherence transfer terms, which require at least two ground states and two excited states, make coherent population oscillations possible}.
Although they do not survive at long times,  they  are the signature  of an important many-level effect.

\section{Numerical Results}
\label{sec:Numerical}

In this section, we show that the dynamics of the four level system display similar regimes to previously studied three-level $V$ and $\Lambda$ systems.
We numerically solve Eq. (1) in the main text by direct diagonalization of the Liouvillian in different parameter regimes.
In all cases, the system is initialized in an equally populated incoherent mixture of ground states $\rho_{g_ig_i}(0)=1/2$.
The addition of a fourth state to this model introduces a large number of new parameters making an exhaustive sweep of the parameter space,  as  has been done for simpler three level models \cite{dodin_quantum_2016, dodin_secular_2018}, intractable.
Instead, we will present a survey of the key regimes of system dynamics and highlight the emergence of interesting new physics.

In previous studies of three level $V$ and $\Lambda$ system models, the dynamics showed two qualitatively different  regimes \cite{dodin_secular_2018,dodin_coherent_2016, dodin_noise-induced_2022, dodin_quantum_2016, tscherbul_long-lived_2014,koyu_long-lived_2018, koyu_long-lived_2022, koyu_steady-state_2021}.
The underdamped regime occurs when the  level splitting $\Delta$ is larger than the decay rate $\gamma$ for the $V$ system or the incoherent pumping rate $r$ for the $\Lambda$ system. 
In this regime, the coherences oscillate before decaying fairly quickly but remain small relative to populations and do not significantly influence population dynamics. 
In contrast, in the overdamped regime where the states are nearly degenerate, coherences are remarkably long-lived, comparable to populations in magnitude and substantially alter population dynamics through interference.

The situation is  slightly more complicated for the four level model we consider here. 
Each of the two manifolds can be in either regime, including mixed regimes where one manifold is overdamped  and the other underdamped. 
This can  be true even if $\Delta_g=\Delta_e$ since the frequency at which dynamics become overdamped is set by $\gamma$ decay rates in the excited  states and $r$ excitation rates in the ground state, which are smaller than the decay  rates by about two orders of magnitude for sunlight excitation in the visible spectrum.
We will begin by  considering the purely underdamped and purely overdamped regimes where the ground and excited states are in the  same regime, showing that the dynamics are consistent  with those of three level models.
Then, we provide a brief description of the dynamics of all density matrix elements in the mixed-damping regime where coherent population oscillations were observed in the main text.

We begin by considering the pure underdamped regime where $\Delta_e/\gamma_{g_ie_j} \gg 1$ and $\Delta_g \gg r_{g_ie_j}$ for all $i$ and $j$. 
In this regime, the timescale of coherence oscillations between the states in each manifold is much shorter than the lifetime  of both states,  allowing for the observation of oscillatory coherences in both manifolds before they decay. 
The coherences oscillate with a frequency similar to the energy splitting between the states and decay at rate $\gamma$ out of the excited state manifold in the $V$-system \cite{tscherbul_long-lived_2014,dodin_coherent_2016} or $\sim r^{-1}$ out of the ground state manifold in a $\Lambda$-system \cite{koyu_long-lived_2022}.
The resulting coherences are relatively small and do not significantly influence population dynamics in  three  level  models.

\begin{figure}[htbp]
	\centering
	\includegraphics[width=\textwidth]{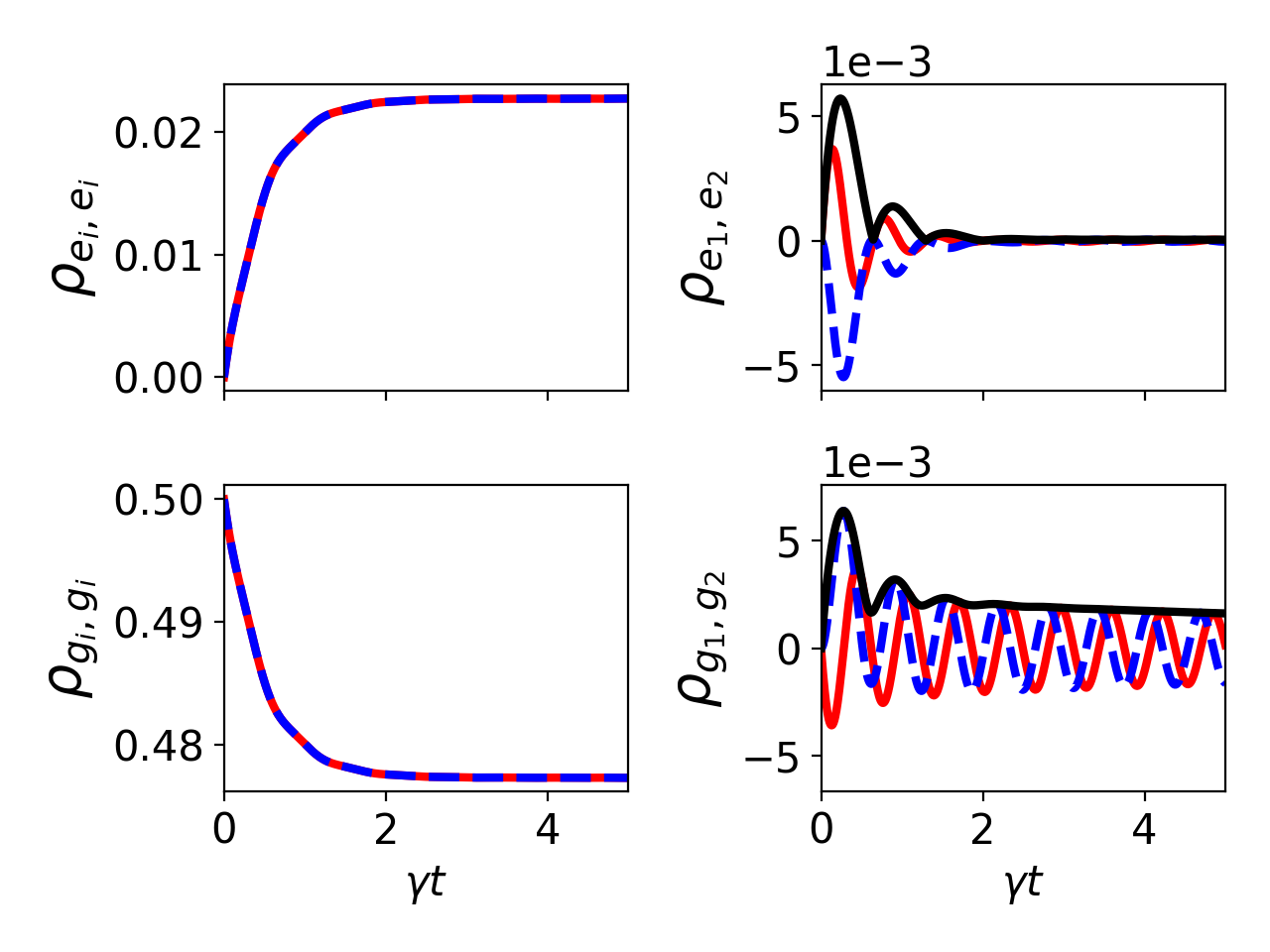}
	\renewcommand{\figurename}{Fig.}
	\caption{Example dynamics of the four-level model in the purely underdamped regime. Populations of state $g_1$ and $e_1$ (red solid) and $g_2$ and $e_2$ (blue dashed) are shown in the left plots. The real (red solid), imaginary (blue dashed) and magnitude (black solid) coherences in each manifold are shown on the right. Time is reported in terms of the mean spontaneous emission rate $\gamma$. All transitions have the same rates $\gamma_{g_ie_j}=\gamma$ and $r_{g_ie_j}=\nbar\gamma$ with $\nbar=0.05$. Ground and excited state manifolds have the same level splitting $\Delta_g=10\gamma=\Delta_e$. All transition dipole moments are taken to be parallel with alignment parameters $p_i=1$.}
\label{fig:UDExample}
\end{figure}

The dynamics in the pure underdamped regime of the four level system, shown in Fig. \ref{fig:UDExample}, appear similar to the $V$ and $\Lambda$ systems.
In both cases, the populations of the ground and excited states evolve in close correspondence to the rate-law equations showing little coherent effects as seen in Fig. \ref{fig:UDExample}.
That is, the population of the excited states grew to a steady state value of $\sim\nbar/2$ at a rate proportional to the $\gamma$ parameters.
The generated coherences are significantly smaller in magnitude than the change in state populations, differing by a factor of $\sim \gamma/\Delta$, similarly to the $V$-system model.
However, the coherences in the excited state decay much faster $\sim \gamma^{-1}$ than those in the ground state $\sim r^{-1}$.
This difference in coherence lifetimes is to be expected from Eq. (1) of the main text since the primary mechanisms of coherence  loss arise from systems leaving an eigenstate.
The small incoherent pumping rate relative to emission thereby translates into a longer ground state  coherence lifetime.

In the overdamped regime, where the spontaneous decay rate is much faster than the period of coherence oscillations, no coherence oscillations are observed in the three level models.
Instead, quasi-stationary coherences are observed that indicate the existence of transient coherent superpositions which survive for a remarkably long time \cite{tscherbul_long-lived_2014, dodin_coherent_2016,koyu_long-lived_2022}.
In the V-system overdamped dynamics occur when $2\sqrt{\Delta^2+(1-p^2)\gamma_1\gamma_2}/(\gamma_1+\gamma_2)\ll1$. 
Motivated by the observation that the rate of decay of ground state coherences in the underdamped regime is given by the pumping rates rather than the spontaneous decay rates we consider the dynamics of the system in the regime $\Delta_i \ll r_j \ll \gamma_k$ for all $i$, $j$  and $k$. 

\begin{figure}[htbp]
	\centering
	\includegraphics[width=\textwidth]{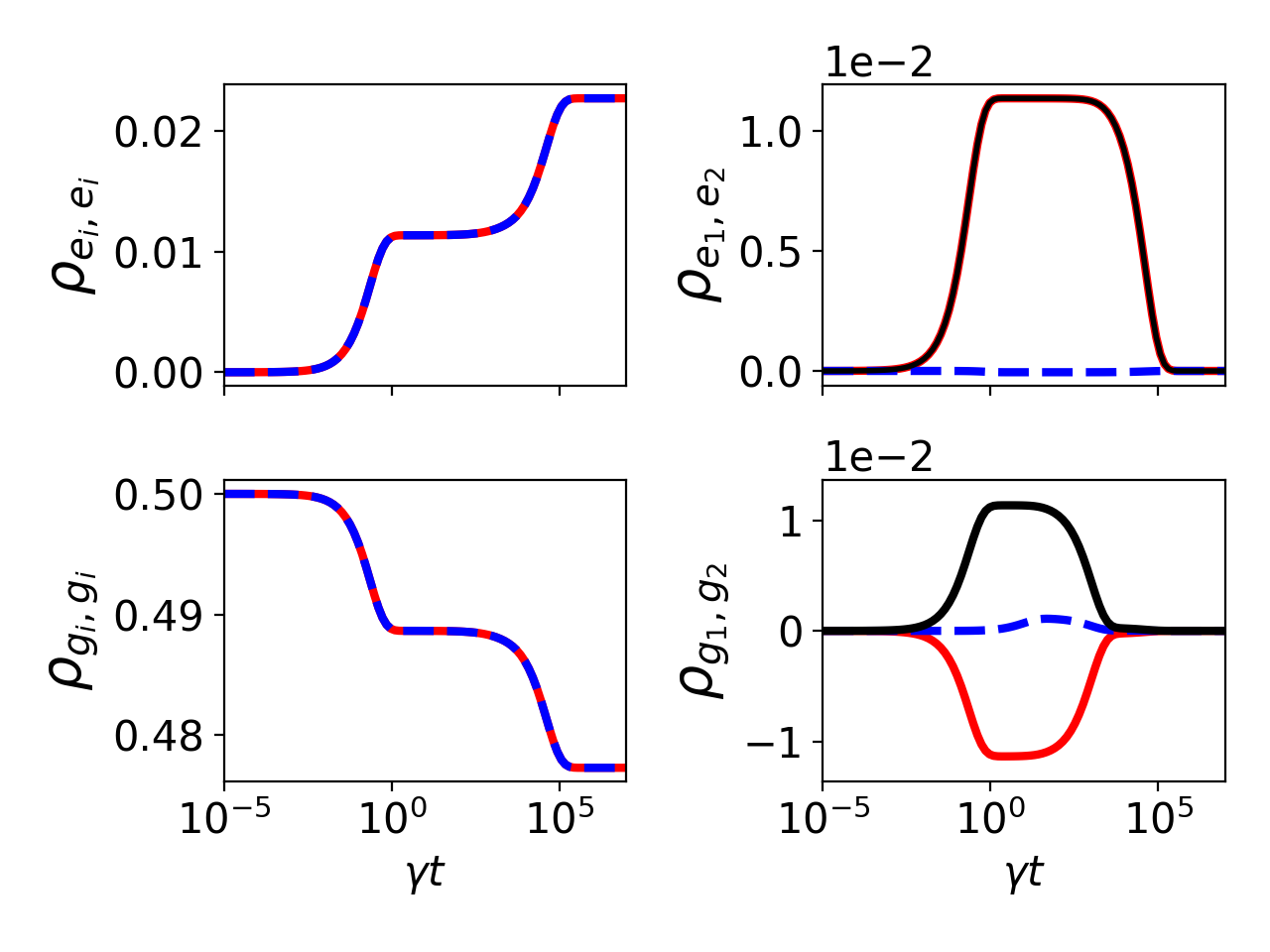}
	\renewcommand{\figurename}{Fig.}
	\caption{Example dynamics of the four-level model in the purely overdamped regime. Populations of state $g_1$ and $e_1$ (red solid) and $g_2$ and $e_2$ (blue dashed) are shown in the left plots. The real (red solid), imaginary (blue dashed) and magnitude (black solid) coherences in each manifold are shown on the right. Time is reported in terms of the mean spontaneous emission rate $\gamma$. All transitions have the same rates $\gamma_{g_ie_j}=\gamma$ and $r_{g_ie_j}=\nbar\gamma$ with $\nbar=0.05$. Ground and excited state manifolds have the same level splitting $\Delta_g=0.01\gamma=\Delta_e$. All transition dipole moments are taken to be parallel with alignment parameters $p_i=1$.}
\label{fig:ODExample}
\end{figure}

An example of the resulting overdamped dynamics showing the transient superposition forming before eventually decaying, at long times, into the thermal state is plotted in Fig. \ref{fig:ODExample}
In contrast to the underdamped regime, the coherences in the ground state are generated 1-2 orders of magnitude slower and survive for 1-2 orders of magnitude shorter time than the excited state.
Three level models predict this discrepancy.
The rise time of the excited state coherence is given  by terms $\sim \gamma^{-1}$ while for ground state coherences it is $\sim r^{-1}$ which is about 1-2 order of magnitude slower.
In contrast, the coherence lifetime is determined by terms of the form $\Delta^2/2\gamma$ for the  excited state  and $\Delta^2/2r$ in the ground state, explaining the difference in lifetimes.

The four level model we consider here admits a new regime that is not possible in the $V$-type and $\Lambda$-type systems  where the states in one manifold are underdamped while those in  the other are overdamped.
This regime is discussed in more detail in the main text.
While this scenario is quite likely to arise in naturally occurring systems, our previous simple models do not provide insight into the consequences of this mixed regime on system dynamics. 
In particular, the ways in which coherence transfer modifies the dynamics from a picture of independent $V$ and $\Lambda$  type transitions remains unclear.

\begin{figure}[htbp]
	\centering
	\includegraphics[width=\textwidth]{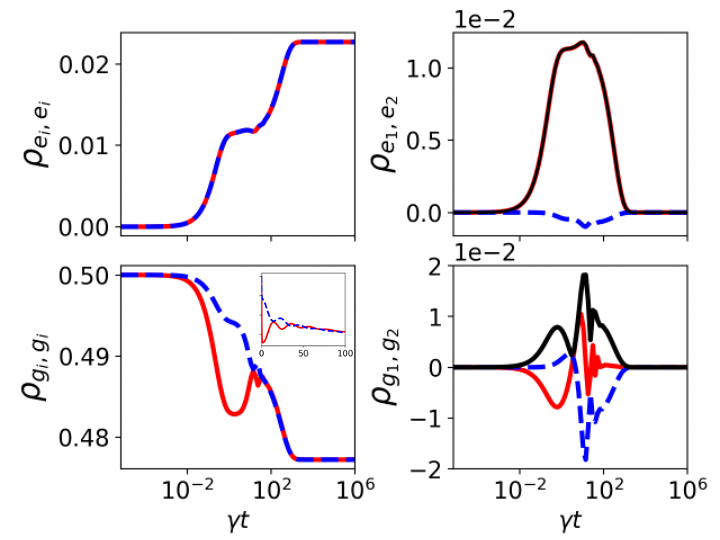}
	\renewcommand{\figurename}{Fig.}
	\caption{Example dynamics of the four-level model in the mixed damping regime. Populations of state $g_1$ and $e_1$ (red solid) and $g_2$ and $e_2$ (blue dashed) are shown in the left plots. The real (red solid), imaginary (blue dashed) and magnitude (black solid) coherences in each manifold are shown on the right. Time is reported in terms of the mean spontaneous emission rate $\gamma$. Transition rates are given by $\gamma_{g_1e_j}=1.5\gamma$,  $\gamma_{g_2e_j}=0.5\gamma$  and $r_{g_ie_j}=\nbar\gamma$ with $\nbar=0.05$. Ground and excited state manifold splittings are $\Delta_g=0.3\gamma$ and $\Delta_e=0.1\gamma$. All transition dipole moments are taken to be parallel with alignment parameters $p_i=1$. The inset highlights the short time dynamics of ground state populations on a linear time axis.}
\label{fig:OUDExample}
\end{figure}

Considering a system in this mixed regime reveals surprising new phenomenology that could not arise in simpler models.
Figure \ref{fig:OUDExample} plots the dynamics of a system where the ground state is in the underdamped regime while the excited state is in the overdamped regime.
While  the excited state shows population dynamics that are not so different from a $V$ system in the overdamped regime, the ground state dynamics are markedly different.
In particular, we see that, rather than entering a quasi-stationary state at intermediate times, the ground state population displays coherent oscillations accompanied by oscillations in the ground state  coherence.
The minimal change in the excited state populations indicates that rather than interference changing the net excitation rate to the excited state, it is altering which ground state excitation originates from.

\begin{figure}
    \centering
    \includegraphics[width=\textwidth]{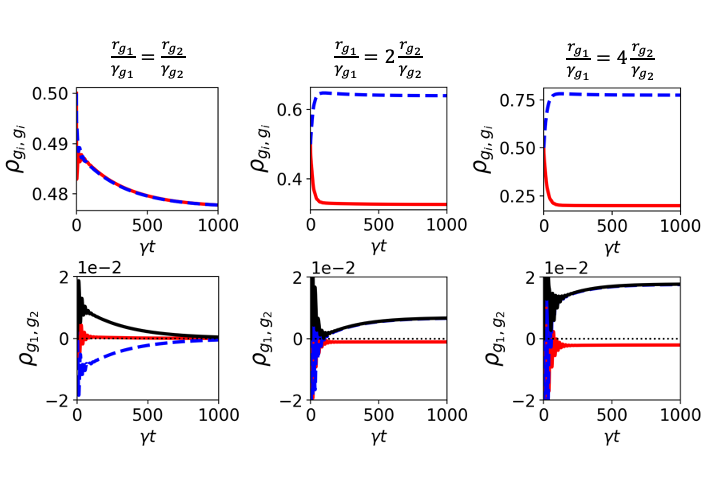}
    \caption{Ground state populations (top) and coherences (bottom) for a non-equilibrium four level system in the mixed damping regime. The population plots show the population of state $g_1$ (red solid lines) and $g_2$ (blue dashed lines) (Note the difference in y-axis scales). The coherrence plots show the real part (red solid lines), imaginary part (blue dashed lines) and magnitude (black solid lines) of the ground state coherences. The dotted black line shows the $\rho_{g_1g_2}=0$ equilibrium state for reference. All parameters as in Fig. 2 of the main text, with $\Delta_e=0.1\gamma$, $\nbar_{g_1}=0.05$ held fixed and $\nbar_{g_2}$ scaled to produce  the appropriate ratios of rates.}
    \label{fig:4LS-NESS}
\end{figure}

Finally, we consider in Fig. \ref{fig:4LS-NESS} the dynamics of a non-equilibrium variant of the model in the mixed damping regime where the two ground states are coupled to baths at  different temperatures.
This leads to a system where the ratio of excitation from and emission out of state $g_1$, $r_{g_1, e_i}/\gamma_{g_1, e_i} = r_{g_1}/\gamma_{g_1}$, is unequal to this ratio for $g_2$, $r_{g_2}/\gamma_{g_2}$.
The difference between these  ratios controls the size of the NESS coherences\cite{koyu_steady-state_2021}, and reflects the temperature difference of the baths coupled to the two ground states, providing a convenient measure for the departure from equilibrium.
As the system departs further from equilibrium, we see two key changes.
First, the population of the ground states becomes increasingly unequal.
This is a classical effect that arises because the two  ground states are coupled to baths at different temperatures.
Second, the magnitude of steady-state coherences increases, reflecting the coherent NESS.
Furthermore, we confirm that the coherent population oscillations are transient responses and do not persist into the NESS.
In all simulations, populations were positive, and remained normalized at all times and coherences satisfied the bound $\rho_{ij}\leq \sqrt{\rho_{ii}\rho_{jj}}$.
Extended numerical simulations to $t=10^{16}\gamma$ showed no change in the plotted steady states.

\section{Pulsed vs Suddenly Turned-On Incoherent Light}
\label{sec:pulse}

In this section, we consider the difference in excitation by pulsed and suddenly turned-on incoherent light.
Specifically, we show that as the pulse duration, $\tau_p$ becomes very short, pulsed incoherent light produces the same type of excitation as a coherent laser pulse and therefore can generate oscillating quantum interference by the same mechanism as laser pulses.
In contrast, suddenly turned on incoherent light remains distinct from laser excitation even in the limit of infinitely fast turn on, requiring the multi-level mechanism we discuss in the main text to generate oscillatory interference.
Both forms of incoherent light are characterized by noisy field phases, leading to a  lack of interference between excitations generated at different times.
The crucial distinction between the two forms of incoherent light lies in \textit{when} the excitations are  produced.
Pulsed incoherent light produces excitations in a short,  specified window of time, while suddenly turned on incoherent light has a specified time where excitations begin but continues to produce excitations at all times in the future.

To illustrate this distinction, consider a classical electric field
\begin{equation}
    \label{eq:semi-classical}
    \bm{E}(t) = \sum_{\bm{k},\lambda}A_{\bm{k},\lambda}(t) \exp(i\phi_{\bm{k},\lambda}(t))\bm{e}_{\bm{k},\lambda},
\end{equation}
where $\bm{e}_{\bm{k},\lambda}$ is the polarization vector, $A_{\bm{k},\lambda}(t)$ is the real-valued amplitude and $\phi_{\bm{k}, \lambda}(t)$ is the complex phase of a field mode with wavevector $\bm{k}$ and polarization vector  $\lambda$.
The amplitude is taken to be a deterministic function that will be used to specify the pulse or turn-on profile of the light, while  the phase will be taken to be a stochastic process that allows us to represent both coherent and incoherent light.

This is a generalization to the usual Fourier-Transform representation of a deterministic electric field which will be useful for our time-domain analysis of incoherent light.
The more familiar representation emerges in the limit that $A_{\bm{k},\lambda}$  becomes a constant (real-valued) Fourier amplitude and $\phi_{\bm{k}, \lambda}(t) = \omega_k t + \phi_0$  where $\omega_k = c k$ is  the  frequency of field mode $k$ determined from the dispersion relation and $\phi_0$ is a constant phase shift.
In general, we do not require that a given field mode is associated with any specific frequency and $\phi_{\bm{k},\lambda}$ is instead a  random variable with correlation time $\tau_c$.
This move to a stochastic representation allows us to describe incoherent light which is characterized by its lack of deterministic phase properties with $\langle\phi_{\bm{k},\lambda}\rangle$ constant in time \footnote{This condition guarantees that  the $\phi$ autocorrelation function decays to a constant value and thus does not induce residual dynamics.}.
However, by abandoning the connection with specific frequencies we will not be able to make use of frequency domain resonance conditions that restrict which field modes couple to each transition.
These resonance conditions must be reached through a time-domain analysis.
We will only be interested in the dynamics averaged over an ensemble of field realizations.
This specification is important since there are infinitely many ensembles of phase functions consistent with incoherent light.
By definition, these all produce the same ensemble averaged field and excitations but the ensemble  of realizations may  be very different  with little  reason to distinguish between one ensemble and another.
As such, the ensemble averaged field and the dynamics it induces are physical quantities that are consistent with the semi-classical limit of the Quantum Master Equations in the main  text while the individual realizations are ill-defined  without additional justification.
Our analysis will not select any one realization and instead will only  require consideration of the correlation time $\tau_c$ of the phase.

In the field representation of Eq. \ref{eq:semi-classical}, an incoherent pulse is characterized by amplitudes, $A_{\bm{k}, \lambda}$, that are small outside of a finite interval $[0, \tau_p]$,  while suddenly turned on light has appreciable intensity on the semi-infinite interval $[0, \infty)$.
When these fields  shine on a material system, they only generate excitations  within these intervals.
If  the material system is initially in state $\ket{i}$, and couples linearly to the field through its dipole moment $\hat{\bm{\mu}}$, then, at time $t$ the field generates an excitation to the state
\begin{equation}
    \label{eq:post-excite}
    \ket{f(t)} = \sum_{\bm{k},\lambda}\sum_{j\neq i}   A_{\bm{k},\lambda}(t) \exp(i\phi_{\bm{k},\lambda}(t))\bm{\mu}_{i,j}\cdot\bm{e}_{\bm{k},\lambda}\ket{j},
\end{equation}
where $\lbrace\ket{j}\rbrace$ is any orthonormal basis containing $\ket{i}$, typically taken to be the energy eigenbasis,  and $\bm{\mu}_{i,j} = \bra{j}\bm\mu\ket{i}$ is the transition dipole moment between states $\ket{i}$ and $\ket{j}$.
If the exciting field is a coherent laser, then excitations generated at different times by the same field mode, $(\bm{k}, \lambda)$, will be related by a well-defined relative phase and can therefore  interfere.
It is precisely this interference between excitations originating at different times  that gives rise to resonance conditions under coherent laser excitation \cite{dodin_noise-induced_2022}.
In contrast, for incoherent light with a finite correlation time  $\tau_c$, only excitations generated within $\tau_c$ of one another have well-defined relative phases.
Outside of this window, different realizations will show different relative phases between excitations generated at different times  (e.g. some  in-phase superpositions and some out-of-phase interference).
Consequently, in  the ensemble average, interference  between excitations occurring at different times will wash out, contributing nothing to dynamics.

Consider the excitations generated in the limit of short pulses.
When the pulse duration $\tau_p\ll\tau_c$, the stochasticity of the incoherent pulse  becomes irrelevant since all excitations are produced within the window $\tau_c$ and can thus interfere. 
Therefore, in the limit of an infinitely short pulse, incoherent light generates  the same  type of excitations as coherent light.
A specific superposition is generated at a specific  time, with its subsequent dynamics  generating oscillatory dynamics in precisely the same manner as an impulsive coherent pulse.
While this limit is extreme, it already demonstrates a fundamental difference between incoherent pulses and suddenly turned on incoherent light.
Even in the limit of an infinitely fast turn on, excitations will continue to be generated outside of the interference window $\tau_c$.
The resulting random phase of subsequently generated excitations will not interfere in the same  way as coherent light subjected  to the same turn on.
Thus, pulsed incoherent light approaches the same short-pulse limit  as coherent light,  while suddenly turned-on incoherent light defines a different limit than suddenly turned-on coherent light.
In this sense, suddenly turned on incoherent light can produce qualitatively new physics while short incoherent pulses simply reproduce the same physics as  coherent pulses.

\bibliography{4LS}

%apsrev4-2.bst 2019-01-14 (MD) hand-edited version of apsrev4-1.bst
%Control: key (0)
%Control: author (8) initials jnrlst
%Control: editor formatted (1) identically to author
%Control: production of article title (-1) disabled
%Control: page (0) single
%Control: year (1) truncated
%Control: production of eprint (0) enabled
\begin{thebibliography}{51}%
\makeatletter
\providecommand \@ifxundefined [1]{%
 \@ifx{#1\undefined}
}%
\providecommand \@ifnum [1]{%
 \ifnum #1\expandafter \@firstoftwo
 \else \expandafter \@secondoftwo
 \fi
}%
\providecommand \@ifx [1]{%
 \ifx #1\expandafter \@firstoftwo
 \else \expandafter \@secondoftwo
 \fi
}%
\providecommand \natexlab [1]{#1}%
\providecommand \enquote  [1]{``#1''}%
\providecommand \bibnamefont  [1]{#1}%
\providecommand \bibfnamefont [1]{#1}%
\providecommand \citenamefont [1]{#1}%
\providecommand \href@noop [0]{\@secondoftwo}%
\providecommand \href [0]{\begingroup \@sanitize@url \@href}%
\providecommand \@href[1]{\@@startlink{#1}\@@href}%
\providecommand \@@href[1]{\endgroup#1\@@endlink}%
\providecommand \@sanitize@url [0]{\catcode `\\12\catcode `\$12\catcode `\&12\catcode `\#12\catcode `\^12\catcode `\_12\catcode `\%12\relax}%
\providecommand \@@startlink[1]{}%
\providecommand \@@endlink[0]{}%
\providecommand \url  [0]{\begingroup\@sanitize@url \@url }%
\providecommand \@url [1]{\endgroup\@href {#1}{\urlprefix }}%
\providecommand \urlprefix  [0]{URL }%
\providecommand \Eprint [0]{\href }%
\providecommand \doibase [0]{https://doi.org/}%
\providecommand \selectlanguage [0]{\@gobble}%
\providecommand \bibinfo  [0]{\@secondoftwo}%
\providecommand \bibfield  [0]{\@secondoftwo}%
\providecommand \translation [1]{[#1]}%
\providecommand \BibitemOpen [0]{}%
\providecommand \bibitemStop [0]{}%
\providecommand \bibitemNoStop [0]{.\EOS\space}%
\providecommand \EOS [0]{\spacefactor3000\relax}%
\providecommand \BibitemShut  [1]{\csname bibitem#1\endcsname}%
\let\auto@bib@innerbib\@empty
%</preamble>
\bibitem [{\citenamefont {Beloy}\ \emph {et~al.}(2006)\citenamefont {Beloy}, \citenamefont {Safronova},\ and\ \citenamefont {Derevianko}}]{beloy_high-accuracy_2006}%
  \BibitemOpen
  \bibfield  {author} {\bibinfo {author} {\bibfnamefont {K.}~\bibnamefont {Beloy}}, \bibinfo {author} {\bibfnamefont {U.~I.}\ \bibnamefont {Safronova}},\ and\ \bibinfo {author} {\bibfnamefont {A.}~\bibnamefont {Derevianko}},\ }\href {https://doi.org/10.1103/PhysRevLett.97.040801} {\bibfield  {journal} {\bibinfo  {journal} {Phys. Rev. Lett.}\ }\textbf {\bibinfo {volume} {97}},\ \bibinfo {pages} {040801} (\bibinfo {year} {2006})}\BibitemShut {NoStop}%
\bibitem [{\citenamefont {Ovsiannikov}\ \emph {et~al.}(2011)\citenamefont {Ovsiannikov}, \citenamefont {Derevianko},\ and\ \citenamefont {Gibble}}]{ovsiannikov_rydberg_2011}%
  \BibitemOpen
  \bibfield  {author} {\bibinfo {author} {\bibfnamefont {V.~D.}\ \bibnamefont {Ovsiannikov}}, \bibinfo {author} {\bibfnamefont {A.}~\bibnamefont {Derevianko}},\ and\ \bibinfo {author} {\bibfnamefont {K.}~\bibnamefont {Gibble}},\ }\href {https://doi.org/10.1103/PhysRevLett.107.093003} {\bibfield  {journal} {\bibinfo  {journal} {Phys. Rev. Lett.}\ }\textbf {\bibinfo {volume} {107}},\ \bibinfo {pages} {093003} (\bibinfo {year} {2011})}\BibitemShut {NoStop}%
\bibitem [{\citenamefont {Safronova}\ \emph {et~al.}(2012)\citenamefont {Safronova}, \citenamefont {Kozlov},\ and\ \citenamefont {Clark}}]{safronova_blackbody_2012}%
  \BibitemOpen
  \bibfield  {author} {\bibinfo {author} {\bibfnamefont {M.~S.}\ \bibnamefont {Safronova}}, \bibinfo {author} {\bibfnamefont {M.~G.}\ \bibnamefont {Kozlov}},\ and\ \bibinfo {author} {\bibfnamefont {C.~W.}\ \bibnamefont {Clark}},\ }\href {https://doi.org/10.1109/TUFFC.2012.2213} {\bibfield  {journal} {\bibinfo  {journal} {IEEE Transactions on Ultrasonics, Ferroelectrics, and Frequency Control}\ }\textbf {\bibinfo {volume} {59}},\ \bibinfo {pages} {439} (\bibinfo {year} {2012})}\BibitemShut {NoStop}%
\bibitem [{\citenamefont {Lisdat}\ \emph {et~al.}(2021)\citenamefont {Lisdat}, \citenamefont {Dörscher}, \citenamefont {Nosske},\ and\ \citenamefont {Sterr}}]{lisdat_blackbody_2021}%
  \BibitemOpen
  \bibfield  {author} {\bibinfo {author} {\bibfnamefont {C.}~\bibnamefont {Lisdat}}, \bibinfo {author} {\bibfnamefont {S.}~\bibnamefont {Dörscher}}, \bibinfo {author} {\bibfnamefont {I.}~\bibnamefont {Nosske}},\ and\ \bibinfo {author} {\bibfnamefont {U.}~\bibnamefont {Sterr}},\ }\href {https://doi.org/10.1103/PhysRevResearch.3.L042036} {\bibfield  {journal} {\bibinfo  {journal} {Phys. Rev. Res.}\ }\textbf {\bibinfo {volume} {3}},\ \bibinfo {pages} {L042036} (\bibinfo {year} {2021})}\BibitemShut {NoStop}%
\bibitem [{\citenamefont {Ajoy}\ \emph {et~al.}(2015)\citenamefont {Ajoy}, \citenamefont {Bissbort}, \citenamefont {Lukin}, \citenamefont {Walsworth},\ and\ \citenamefont {Cappellaro}}]{ajoy_atomic-scale_2015}%
  \BibitemOpen
  \bibfield  {author} {\bibinfo {author} {\bibfnamefont {A.}~\bibnamefont {Ajoy}}, \bibinfo {author} {\bibfnamefont {U.}~\bibnamefont {Bissbort}}, \bibinfo {author} {\bibfnamefont {M.}~\bibnamefont {Lukin}}, \bibinfo {author} {\bibfnamefont {R.}~\bibnamefont {Walsworth}},\ and\ \bibinfo {author} {\bibfnamefont {P.}~\bibnamefont {Cappellaro}},\ }\href {https://doi.org/10.1103/PhysRevX.5.011001} {\bibfield  {journal} {\bibinfo  {journal} {Phys. Rev. X}\ }\textbf {\bibinfo {volume} {5}},\ \bibinfo {pages} {011001} (\bibinfo {year} {2015})}\BibitemShut {NoStop}%
\bibitem [{\citenamefont {Stanwix}\ \emph {et~al.}(2010)\citenamefont {Stanwix}, \citenamefont {Pham}, \citenamefont {Maze}, \citenamefont {Le~Sage}, \citenamefont {Yeung}, \citenamefont {Cappellaro}, \citenamefont {Hemmer}, \citenamefont {Yacoby}, \citenamefont {Lukin},\ and\ \citenamefont {Walsworth}}]{stanwix_coherence_2010}%
  \BibitemOpen
  \bibfield  {author} {\bibinfo {author} {\bibfnamefont {P.~L.}\ \bibnamefont {Stanwix}}, \bibinfo {author} {\bibfnamefont {L.~M.}\ \bibnamefont {Pham}}, \bibinfo {author} {\bibfnamefont {J.~R.}\ \bibnamefont {Maze}}, \bibinfo {author} {\bibfnamefont {D.}~\bibnamefont {Le~Sage}}, \bibinfo {author} {\bibfnamefont {T.~K.}\ \bibnamefont {Yeung}}, \bibinfo {author} {\bibfnamefont {P.}~\bibnamefont {Cappellaro}}, \bibinfo {author} {\bibfnamefont {P.~R.}\ \bibnamefont {Hemmer}}, \bibinfo {author} {\bibfnamefont {A.}~\bibnamefont {Yacoby}}, \bibinfo {author} {\bibfnamefont {M.~D.}\ \bibnamefont {Lukin}},\ and\ \bibinfo {author} {\bibfnamefont {R.~L.}\ \bibnamefont {Walsworth}},\ }\href@noop {} {\bibfield  {journal} {\bibinfo  {journal} {Physical Review B}\ }\textbf {\bibinfo {volume} {82}},\ \bibinfo {pages} {201201} (\bibinfo {year} {2010})}\BibitemShut {NoStop}%
\bibitem [{\citenamefont {Bar-Gill}\ \emph {et~al.}(2012)\citenamefont {Bar-Gill}, \citenamefont {Pham}, \citenamefont {Belthangady}, \citenamefont {Le~Sage}, \citenamefont {Cappellaro}, \citenamefont {Maze}, \citenamefont {Lukin}, \citenamefont {Yacoby},\ and\ \citenamefont {Walsworth}}]{bar-gill_suppression_2012}%
  \BibitemOpen
  \bibfield  {author} {\bibinfo {author} {\bibfnamefont {N.}~\bibnamefont {Bar-Gill}}, \bibinfo {author} {\bibfnamefont {L.~M.}\ \bibnamefont {Pham}}, \bibinfo {author} {\bibfnamefont {C.}~\bibnamefont {Belthangady}}, \bibinfo {author} {\bibfnamefont {D.}~\bibnamefont {Le~Sage}}, \bibinfo {author} {\bibfnamefont {P.}~\bibnamefont {Cappellaro}}, \bibinfo {author} {\bibfnamefont {J.~R.}\ \bibnamefont {Maze}}, \bibinfo {author} {\bibfnamefont {M.~D.}\ \bibnamefont {Lukin}}, \bibinfo {author} {\bibfnamefont {A.}~\bibnamefont {Yacoby}},\ and\ \bibinfo {author} {\bibfnamefont {R.}~\bibnamefont {Walsworth}},\ }\href {https://doi.org/10.1038/ncomms1856} {\bibfield  {journal} {\bibinfo  {journal} {Nat Commun}\ }\textbf {\bibinfo {volume} {3}},\ \bibinfo {pages} {858} (\bibinfo {year} {2012})}\BibitemShut {NoStop}%
\bibitem [{\citenamefont {Degen}\ \emph {et~al.}(2017)\citenamefont {Degen}, \citenamefont {Reinhard},\ and\ \citenamefont {Cappellaro}}]{degen_quantum_2017}%
  \BibitemOpen
  \bibfield  {author} {\bibinfo {author} {\bibfnamefont {C.}~\bibnamefont {Degen}}, \bibinfo {author} {\bibfnamefont {F.}~\bibnamefont {Reinhard}},\ and\ \bibinfo {author} {\bibfnamefont {P.}~\bibnamefont {Cappellaro}},\ }\href {https://doi.org/10.1103/RevModPhys.89.035002} {\bibfield  {journal} {\bibinfo  {journal} {Rev. Mod. Phys.}\ }\textbf {\bibinfo {volume} {89}},\ \bibinfo {pages} {035002} (\bibinfo {year} {2017})}\BibitemShut {NoStop}%
\bibitem [{\citenamefont {Dolde}\ \emph {et~al.}(2011)\citenamefont {Dolde}, \citenamefont {Fedder}, \citenamefont {Doherty}, \citenamefont {Nöbauer}, \citenamefont {Rempp}, \citenamefont {Balasubramanian}, \citenamefont {Wolf}, \citenamefont {Reinhard}, \citenamefont {Hollenberg}, \citenamefont {Jelezko},\ and\ \citenamefont {Wrachtrup}}]{dolde_electric-field_2011}%
  \BibitemOpen
  \bibfield  {author} {\bibinfo {author} {\bibfnamefont {F.}~\bibnamefont {Dolde}}, \bibinfo {author} {\bibfnamefont {H.}~\bibnamefont {Fedder}}, \bibinfo {author} {\bibfnamefont {M.~W.}\ \bibnamefont {Doherty}}, \bibinfo {author} {\bibfnamefont {T.}~\bibnamefont {Nöbauer}}, \bibinfo {author} {\bibfnamefont {F.}~\bibnamefont {Rempp}}, \bibinfo {author} {\bibfnamefont {G.}~\bibnamefont {Balasubramanian}}, \bibinfo {author} {\bibfnamefont {T.}~\bibnamefont {Wolf}}, \bibinfo {author} {\bibfnamefont {F.}~\bibnamefont {Reinhard}}, \bibinfo {author} {\bibfnamefont {L.~C.~L.}\ \bibnamefont {Hollenberg}}, \bibinfo {author} {\bibfnamefont {F.}~\bibnamefont {Jelezko}},\ and\ \bibinfo {author} {\bibfnamefont {J.}~\bibnamefont {Wrachtrup}},\ }\href {https://doi.org/10.1038/nphys1969} {\bibfield  {journal} {\bibinfo  {journal} {Nature Phys}\ }\textbf {\bibinfo {volume} {7}},\ \bibinfo {pages} {459} (\bibinfo {year} {2011})}\BibitemShut {NoStop}%
\bibitem [{\citenamefont {Doherty}\ \emph {et~al.}(2013)\citenamefont {Doherty}, \citenamefont {Manson}, \citenamefont {Delaney}, \citenamefont {Jelezko}, \citenamefont {Wrachtrup},\ and\ \citenamefont {Hollenberg}}]{doherty_nitrogen-vacancy_2013}%
  \BibitemOpen
  \bibfield  {author} {\bibinfo {author} {\bibfnamefont {M.~W.}\ \bibnamefont {Doherty}}, \bibinfo {author} {\bibfnamefont {N.~B.}\ \bibnamefont {Manson}}, \bibinfo {author} {\bibfnamefont {P.}~\bibnamefont {Delaney}}, \bibinfo {author} {\bibfnamefont {F.}~\bibnamefont {Jelezko}}, \bibinfo {author} {\bibfnamefont {J.}~\bibnamefont {Wrachtrup}},\ and\ \bibinfo {author} {\bibfnamefont {L.~C.}\ \bibnamefont {Hollenberg}},\ }\href@noop {} {\bibfield  {journal} {\bibinfo  {journal} {Physics Reports}\ }\textbf {\bibinfo {volume} {528}},\ \bibinfo {pages} {1} (\bibinfo {year} {2013})}\BibitemShut {NoStop}%
\bibitem [{\citenamefont {Neumann}\ \emph {et~al.}(2013)\citenamefont {Neumann}, \citenamefont {Jakobi}, \citenamefont {Dolde}, \citenamefont {Burk}, \citenamefont {Reuter}, \citenamefont {Waldherr}, \citenamefont {Honert}, \citenamefont {Wolf}, \citenamefont {Brunner}, \citenamefont {Shim}, \citenamefont {Suter}, \citenamefont {Sumiya}, \citenamefont {Isoya},\ and\ \citenamefont {Wrachtrup}}]{neumann_high-precision_2013}%
  \BibitemOpen
  \bibfield  {author} {\bibinfo {author} {\bibfnamefont {P.}~\bibnamefont {Neumann}}, \bibinfo {author} {\bibfnamefont {I.}~\bibnamefont {Jakobi}}, \bibinfo {author} {\bibfnamefont {F.}~\bibnamefont {Dolde}}, \bibinfo {author} {\bibfnamefont {C.}~\bibnamefont {Burk}}, \bibinfo {author} {\bibfnamefont {R.}~\bibnamefont {Reuter}}, \bibinfo {author} {\bibfnamefont {G.}~\bibnamefont {Waldherr}}, \bibinfo {author} {\bibfnamefont {J.}~\bibnamefont {Honert}}, \bibinfo {author} {\bibfnamefont {T.}~\bibnamefont {Wolf}}, \bibinfo {author} {\bibfnamefont {A.}~\bibnamefont {Brunner}}, \bibinfo {author} {\bibfnamefont {J.~H.}\ \bibnamefont {Shim}}, \bibinfo {author} {\bibfnamefont {D.}~\bibnamefont {Suter}}, \bibinfo {author} {\bibfnamefont {H.}~\bibnamefont {Sumiya}}, \bibinfo {author} {\bibfnamefont {J.}~\bibnamefont {Isoya}},\ and\ \bibinfo {author} {\bibfnamefont {J.}~\bibnamefont {Wrachtrup}},\ }\href {https://doi.org/10.1021/nl401216y} {\bibfield  {journal} {\bibinfo  {journal} {Nano Lett.}\ }\textbf {\bibinfo
  {volume} {13}},\ \bibinfo {pages} {2738} (\bibinfo {year} {2013})}\BibitemShut {NoStop}%
\bibitem [{\citenamefont {Zhang}\ \emph {et~al.}(2022)\citenamefont {Zhang}, \citenamefont {Dasari}, \citenamefont {Widmann}, \citenamefont {Meinel}, \citenamefont {Vorobyov}, \citenamefont {Kapitanova}, \citenamefont {Nenasheva}, \citenamefont {Nakamura}, \citenamefont {Sumiya}, \citenamefont {Onoda}, \citenamefont {Isoya},\ and\ \citenamefont {Wrachtrup}}]{zhang_quantum-assisted_2022}%
  \BibitemOpen
  \bibfield  {author} {\bibinfo {author} {\bibfnamefont {C.}~\bibnamefont {Zhang}}, \bibinfo {author} {\bibfnamefont {D.}~\bibnamefont {Dasari}}, \bibinfo {author} {\bibfnamefont {M.}~\bibnamefont {Widmann}}, \bibinfo {author} {\bibfnamefont {J.}~\bibnamefont {Meinel}}, \bibinfo {author} {\bibfnamefont {V.}~\bibnamefont {Vorobyov}}, \bibinfo {author} {\bibfnamefont {P.}~\bibnamefont {Kapitanova}}, \bibinfo {author} {\bibfnamefont {E.}~\bibnamefont {Nenasheva}}, \bibinfo {author} {\bibfnamefont {K.}~\bibnamefont {Nakamura}}, \bibinfo {author} {\bibfnamefont {H.}~\bibnamefont {Sumiya}}, \bibinfo {author} {\bibfnamefont {S.}~\bibnamefont {Onoda}}, \bibinfo {author} {\bibfnamefont {J.}~\bibnamefont {Isoya}},\ and\ \bibinfo {author} {\bibfnamefont {J.}~\bibnamefont {Wrachtrup}},\ }\href {https://doi.org/10.1038/s41467-022-32150-1} {\bibfield  {journal} {\bibinfo  {journal} {Nat Commun}\ }\textbf {\bibinfo {volume} {13}},\ \bibinfo {pages} {4637} (\bibinfo {year} {2022})}\BibitemShut {NoStop}%
\bibitem [{\citenamefont {Nielsen}\ and\ \citenamefont {Chuang}(2010)}]{nielsen_quantum_2010}%
  \BibitemOpen
  \bibfield  {author} {\bibinfo {author} {\bibfnamefont {M.~A.}\ \bibnamefont {Nielsen}}\ and\ \bibinfo {author} {\bibfnamefont {I.~L.}\ \bibnamefont {Chuang}},\ }\href@noop {} {\emph {\bibinfo {title} {Quantum computation and quantum information}}}\ (\bibinfo  {publisher} {Cambridge University Press},\ \bibinfo {year} {2010})\BibitemShut {NoStop}%
\bibitem [{\citenamefont {Monroe}(2002)}]{monroe_quantum_2002}%
  \BibitemOpen
  \bibfield  {author} {\bibinfo {author} {\bibfnamefont {C.}~\bibnamefont {Monroe}},\ }\href {https://doi.org/10.1038/416238a} {\bibfield  {journal} {\bibinfo  {journal} {Nature}\ }\textbf {\bibinfo {volume} {416}},\ \bibinfo {pages} {238} (\bibinfo {year} {2002})}\BibitemShut {NoStop}%
\bibitem [{\citenamefont {Preskill}(2018)}]{preskill_quantum_2018}%
  \BibitemOpen
  \bibfield  {author} {\bibinfo {author} {\bibfnamefont {J.}~\bibnamefont {Preskill}},\ }\href {https://doi.org/10.22331/q-2018-08-06-79} {\bibfield  {journal} {\bibinfo  {journal} {Quantum}\ }\textbf {\bibinfo {volume} {2}},\ \bibinfo {pages} {79} (\bibinfo {year} {2018})}\BibitemShut {NoStop}%
\bibitem [{\citenamefont {Resch}\ and\ \citenamefont {Karpuzcu}(2021)}]{resch_benchmarking_2021}%
  \BibitemOpen
  \bibfield  {author} {\bibinfo {author} {\bibfnamefont {S.}~\bibnamefont {Resch}}\ and\ \bibinfo {author} {\bibfnamefont {U.~R.}\ \bibnamefont {Karpuzcu}},\ }\href {https://doi.org/10.1145/3464420} {\bibfield  {journal} {\bibinfo  {journal} {ACM Comput. Surv.}\ }\textbf {\bibinfo {volume} {54}},\ \bibinfo {pages} {142:1} (\bibinfo {year} {2021})}\BibitemShut {NoStop}%
\bibitem [{\citenamefont {Dawson}\ \emph {et~al.}(2006)\citenamefont {Dawson}, \citenamefont {Haselgrove},\ and\ \citenamefont {Nielsen}}]{dawson_noise_2006}%
  \BibitemOpen
  \bibfield  {author} {\bibinfo {author} {\bibfnamefont {C.~M.}\ \bibnamefont {Dawson}}, \bibinfo {author} {\bibfnamefont {H.~L.}\ \bibnamefont {Haselgrove}},\ and\ \bibinfo {author} {\bibfnamefont {M.~A.}\ \bibnamefont {Nielsen}},\ }\href {https://doi.org/10.1103/PhysRevLett.96.020501} {\bibfield  {journal} {\bibinfo  {journal} {Phys. Rev. Lett.}\ }\textbf {\bibinfo {volume} {96}},\ \bibinfo {pages} {020501} (\bibinfo {year} {2006})}\BibitemShut {NoStop}%
\bibitem [{\citenamefont {Bharti}\ \emph {et~al.}(2022)\citenamefont {Bharti}, \citenamefont {Cervera-Lierta}, \citenamefont {Kyaw}, \citenamefont {Haug}, \citenamefont {Alperin-Lea}, \citenamefont {Anand}, \citenamefont {Degroote}, \citenamefont {Heimonen}, \citenamefont {Kottmann}, \citenamefont {Menke}, \citenamefont {Mok}, \citenamefont {Sim}, \citenamefont {Kwek},\ and\ \citenamefont {Aspuru-Guzik}}]{bharti_noisy_2022}%
  \BibitemOpen
  \bibfield  {author} {\bibinfo {author} {\bibfnamefont {K.}~\bibnamefont {Bharti}}, \bibinfo {author} {\bibfnamefont {A.}~\bibnamefont {Cervera-Lierta}}, \bibinfo {author} {\bibfnamefont {T.~H.}\ \bibnamefont {Kyaw}}, \bibinfo {author} {\bibfnamefont {T.}~\bibnamefont {Haug}}, \bibinfo {author} {\bibfnamefont {S.}~\bibnamefont {Alperin-Lea}}, \bibinfo {author} {\bibfnamefont {A.}~\bibnamefont {Anand}}, \bibinfo {author} {\bibfnamefont {M.}~\bibnamefont {Degroote}}, \bibinfo {author} {\bibfnamefont {H.}~\bibnamefont {Heimonen}}, \bibinfo {author} {\bibfnamefont {J.~S.}\ \bibnamefont {Kottmann}}, \bibinfo {author} {\bibfnamefont {T.}~\bibnamefont {Menke}}, \bibinfo {author} {\bibfnamefont {W.-K.}\ \bibnamefont {Mok}}, \bibinfo {author} {\bibfnamefont {S.}~\bibnamefont {Sim}}, \bibinfo {author} {\bibfnamefont {L.-C.}\ \bibnamefont {Kwek}},\ and\ \bibinfo {author} {\bibfnamefont {A.}~\bibnamefont {Aspuru-Guzik}},\ }\href {https://doi.org/10.1103/RevModPhys.94.015004} {\bibfield  {journal} {\bibinfo  {journal}
  {Rev. Mod. Phys.}\ }\textbf {\bibinfo {volume} {94}},\ \bibinfo {pages} {015004} (\bibinfo {year} {2022})}\BibitemShut {NoStop}%
\bibitem [{\citenamefont {Dodin}\ and\ \citenamefont {Brumer}(2022)}]{dodin_noise-induced_2022}%
  \BibitemOpen
  \bibfield  {author} {\bibinfo {author} {\bibfnamefont {A.}~\bibnamefont {Dodin}}\ and\ \bibinfo {author} {\bibfnamefont {P.}~\bibnamefont {Brumer}},\ }\href {https://doi.org/10.1088/1361-6455/ac3e77} {\bibfield  {journal} {\bibinfo  {journal} {J. Phys. B: At. Mol. Opt. Phys.}\ }\textbf {\bibinfo {volume} {54}},\ \bibinfo {pages} {223001} (\bibinfo {year} {2022})}\BibitemShut {NoStop}%
\bibitem [{\citenamefont {Dodin}\ \emph {et~al.}(2018)\citenamefont {Dodin}, \citenamefont {Tscherbul}, \citenamefont {Alicki}, \citenamefont {Vutha},\ and\ \citenamefont {Brumer}}]{dodin_secular_2018}%
  \BibitemOpen
  \bibfield  {author} {\bibinfo {author} {\bibfnamefont {A.}~\bibnamefont {Dodin}}, \bibinfo {author} {\bibfnamefont {T.}~\bibnamefont {Tscherbul}}, \bibinfo {author} {\bibfnamefont {R.}~\bibnamefont {Alicki}}, \bibinfo {author} {\bibfnamefont {A.}~\bibnamefont {Vutha}},\ and\ \bibinfo {author} {\bibfnamefont {P.}~\bibnamefont {Brumer}},\ }\href@noop {} {\bibfield  {journal} {\bibinfo  {journal} {Physical Review A}\ }\textbf {\bibinfo {volume} {97}},\ \bibinfo {pages} {013421} (\bibinfo {year} {2018})}\BibitemShut {NoStop}%
\bibitem [{\citenamefont {Koyu}\ \emph {et~al.}(2021)\citenamefont {Koyu}, \citenamefont {Dodin}, \citenamefont {Brumer},\ and\ \citenamefont {Tscherbul}}]{koyu_steady-state_2021}%
  \BibitemOpen
  \bibfield  {author} {\bibinfo {author} {\bibfnamefont {S.}~\bibnamefont {Koyu}}, \bibinfo {author} {\bibfnamefont {A.}~\bibnamefont {Dodin}}, \bibinfo {author} {\bibfnamefont {P.}~\bibnamefont {Brumer}},\ and\ \bibinfo {author} {\bibfnamefont {T.~V.}\ \bibnamefont {Tscherbul}},\ }\href {https://doi.org/10.1103/PhysRevResearch.3.013295} {\bibfield  {journal} {\bibinfo  {journal} {Phys. Rev. Res.}\ }\textbf {\bibinfo {volume} {3}},\ \bibinfo {pages} {013295} (\bibinfo {year} {2021})}\BibitemShut {NoStop}%
\bibitem [{\citenamefont {Koyu}\ and\ \citenamefont {Tscherbul}(2022)}]{koyu_long-lived_2022}%
  \BibitemOpen
  \bibfield  {author} {\bibinfo {author} {\bibfnamefont {S.}~\bibnamefont {Koyu}}\ and\ \bibinfo {author} {\bibfnamefont {T.~V.}\ \bibnamefont {Tscherbul}},\ }\href {https://doi.org/10.1063/5.0102808} {\bibfield  {journal} {\bibinfo  {journal} {J. Chem. Phys.}\ }\textbf {\bibinfo {volume} {157}},\ \bibinfo {pages} {124302} (\bibinfo {year} {2022})}\BibitemShut {NoStop}%
\bibitem [{\citenamefont {Scully}\ \emph {et~al.}(2011)\citenamefont {Scully}, \citenamefont {Chapin}, \citenamefont {Dorfman}, \citenamefont {Kim},\ and\ \citenamefont {Svidzinsky}}]{scully_quantum_2011}%
  \BibitemOpen
  \bibfield  {author} {\bibinfo {author} {\bibfnamefont {M.~O.}\ \bibnamefont {Scully}}, \bibinfo {author} {\bibfnamefont {K.~R.}\ \bibnamefont {Chapin}}, \bibinfo {author} {\bibfnamefont {K.~E.}\ \bibnamefont {Dorfman}}, \bibinfo {author} {\bibfnamefont {M.~B.}\ \bibnamefont {Kim}},\ and\ \bibinfo {author} {\bibfnamefont {A.}~\bibnamefont {Svidzinsky}},\ }\href {https://doi.org/10.1073/pnas.1110234108} {\bibfield  {journal} {\bibinfo  {journal} {Proc. Natl. Acad. Sci. USA}\ }\textbf {\bibinfo {volume} {108}},\ \bibinfo {pages} {15097} (\bibinfo {year} {2011})}\BibitemShut {NoStop}%
\bibitem [{\citenamefont {Gelbwaser-Klimovsky}\ \emph {et~al.}(2015)\citenamefont {Gelbwaser-Klimovsky}, \citenamefont {Niedenzu}, \citenamefont {Brumer},\ and\ \citenamefont {Kurizki}}]{gelbwaser-klimovsky_power_2015}%
  \BibitemOpen
  \bibfield  {author} {\bibinfo {author} {\bibfnamefont {D.}~\bibnamefont {Gelbwaser-Klimovsky}}, \bibinfo {author} {\bibfnamefont {W.}~\bibnamefont {Niedenzu}}, \bibinfo {author} {\bibfnamefont {P.}~\bibnamefont {Brumer}},\ and\ \bibinfo {author} {\bibfnamefont {G.}~\bibnamefont {Kurizki}},\ }\bibfield  {journal} {\bibinfo  {journal} {Sci Rep}\ }\textbf {\bibinfo {volume} {5}},\ \href {https://doi.org/10.1038/srep14413} {10.1038/srep14413} (\bibinfo {year} {2015})\BibitemShut {NoStop}%
\bibitem [{\citenamefont {Tscherbul}\ and\ \citenamefont {Brumer}(2014{\natexlab{a}})}]{tscherbul_excitation_2014}%
  \BibitemOpen
  \bibfield  {author} {\bibinfo {author} {\bibfnamefont {T.~V.}\ \bibnamefont {Tscherbul}}\ and\ \bibinfo {author} {\bibfnamefont {P.}~\bibnamefont {Brumer}},\ }\href {https://doi.org/10.1021/jp501700t} {\bibfield  {journal} {\bibinfo  {journal} {J. Phys. Chem. A}\ }\textbf {\bibinfo {volume} {118}},\ \bibinfo {pages} {3100} (\bibinfo {year} {2014}{\natexlab{a}})}\BibitemShut {NoStop}%
\bibitem [{\citenamefont {Tscherbul}\ and\ \citenamefont {Brumer}(2015{\natexlab{a}})}]{tscherbul_quantum_2015}%
  \BibitemOpen
  \bibfield  {author} {\bibinfo {author} {\bibfnamefont {T.~V.}\ \bibnamefont {Tscherbul}}\ and\ \bibinfo {author} {\bibfnamefont {P.}~\bibnamefont {Brumer}},\ }\href {https://doi.org/10.1039/C5CP01388G} {\bibfield  {journal} {\bibinfo  {journal} {Phys. Chem. Chem. Phys.}\ }\textbf {\bibinfo {volume} {17}},\ \bibinfo {pages} {30904} (\bibinfo {year} {2015}{\natexlab{a}})}\BibitemShut {NoStop}%
\bibitem [{\citenamefont {Dodin}\ and\ \citenamefont {Brumer}(2019)}]{dodin_light-induced_2019}%
  \BibitemOpen
  \bibfield  {author} {\bibinfo {author} {\bibfnamefont {A.}~\bibnamefont {Dodin}}\ and\ \bibinfo {author} {\bibfnamefont {P.}~\bibnamefont {Brumer}},\ }\href {https://doi.org/10.1063/1.5092981} {\bibfield  {journal} {\bibinfo  {journal} {J. Chem. Phys.}\ }\textbf {\bibinfo {volume} {150}},\ \bibinfo {pages} {184304} (\bibinfo {year} {2019})}\BibitemShut {NoStop}%
\bibitem [{\citenamefont {Tscherbul}\ and\ \citenamefont {Brumer}(2015{\natexlab{b}})}]{tscherbul_partial_2015}%
  \BibitemOpen
  \bibfield  {author} {\bibinfo {author} {\bibfnamefont {T.~V.}\ \bibnamefont {Tscherbul}}\ and\ \bibinfo {author} {\bibfnamefont {P.}~\bibnamefont {Brumer}},\ }\href {https://doi.org/10.1063/1.4908130} {\bibfield  {journal} {\bibinfo  {journal} {J. Chem. Phys.}\ }\textbf {\bibinfo {volume} {142}},\ \bibinfo {pages} {104107} (\bibinfo {year} {2015}{\natexlab{b}})}\BibitemShut {NoStop}%
\bibitem [{\citenamefont {Svidzinsky}\ \emph {et~al.}(2011)\citenamefont {Svidzinsky}, \citenamefont {Dorfman},\ and\ \citenamefont {Scully}}]{svidzinsky_enhancing_2011}%
  \BibitemOpen
  \bibfield  {author} {\bibinfo {author} {\bibfnamefont {A.~A.}\ \bibnamefont {Svidzinsky}}, \bibinfo {author} {\bibfnamefont {K.~E.}\ \bibnamefont {Dorfman}},\ and\ \bibinfo {author} {\bibfnamefont {M.~O.}\ \bibnamefont {Scully}},\ }\href {https://doi.org/10.1103/PhysRevA.84.053818} {\bibfield  {journal} {\bibinfo  {journal} {Phys. Rev. A}\ }\textbf {\bibinfo {volume} {84}},\ \bibinfo {pages} {053818} (\bibinfo {year} {2011})}\BibitemShut {NoStop}%
\bibitem [{\citenamefont {Jung}\ and\ \citenamefont {Brumer}(2020)}]{jung_energy_2020}%
  \BibitemOpen
  \bibfield  {author} {\bibinfo {author} {\bibfnamefont {K.~A.}\ \bibnamefont {Jung}}\ and\ \bibinfo {author} {\bibfnamefont {P.}~\bibnamefont {Brumer}},\ }\href {https://doi.org/10.1063/5.0020576} {\bibfield  {journal} {\bibinfo  {journal} {J. Chem. Phys.}\ }\textbf {\bibinfo {volume} {153}},\ \bibinfo {pages} {114102} (\bibinfo {year} {2020})}\BibitemShut {NoStop}%
\bibitem [{\citenamefont {Joubert-Doriol}\ \emph {et~al.}(2023)\citenamefont {Joubert-Doriol}, \citenamefont {Jung}, \citenamefont {Izmaylov},\ and\ \citenamefont {Brumer}}]{joubert-doriol_quantum_2023}%
  \BibitemOpen
  \bibfield  {author} {\bibinfo {author} {\bibfnamefont {L.}~\bibnamefont {Joubert-Doriol}}, \bibinfo {author} {\bibfnamefont {K.~A.}\ \bibnamefont {Jung}}, \bibinfo {author} {\bibfnamefont {A.~F.}\ \bibnamefont {Izmaylov}},\ and\ \bibinfo {author} {\bibfnamefont {P.}~\bibnamefont {Brumer}},\ }\href {https://doi.org/10.1021/acs.jctc.2c00987} {\bibfield  {journal} {\bibinfo  {journal} {J. Chem. Theory Comput.}\ }\textbf {\bibinfo {volume} {19}},\ \bibinfo {pages} {1130} (\bibinfo {year} {2023})}\BibitemShut {NoStop}%
\bibitem [{\citenamefont {Ivander}\ \emph {et~al.}(2022)\citenamefont {Ivander}, \citenamefont {Anto-Sztrikacs},\ and\ \citenamefont {Segal}}]{ivander_quantum_2022}%
  \BibitemOpen
  \bibfield  {author} {\bibinfo {author} {\bibfnamefont {F.}~\bibnamefont {Ivander}}, \bibinfo {author} {\bibfnamefont {N.}~\bibnamefont {Anto-Sztrikacs}},\ and\ \bibinfo {author} {\bibfnamefont {D.}~\bibnamefont {Segal}},\ }\href {https://doi.org/10.1088/1367-2630/ac9498} {\bibfield  {journal} {\bibinfo  {journal} {New J. Phys.}\ }\textbf {\bibinfo {volume} {24}},\ \bibinfo {pages} {103010} (\bibinfo {year} {2022})}\BibitemShut {NoStop}%
\bibitem [{\citenamefont {Grinev}\ \emph {et~al.}(2015)\citenamefont {Grinev}, \citenamefont {Shapiro},\ and\ \citenamefont {Brumer}}]{grinev_coherent_2015}%
  \BibitemOpen
  \bibfield  {author} {\bibinfo {author} {\bibfnamefont {T.}~\bibnamefont {Grinev}}, \bibinfo {author} {\bibfnamefont {M.}~\bibnamefont {Shapiro}},\ and\ \bibinfo {author} {\bibfnamefont {P.}~\bibnamefont {Brumer}},\ }\href {https://doi.org/10.1088/0953-4075/48/17/174004} {\bibfield  {journal} {\bibinfo  {journal} {J. Phys. B: At. Mol. Opt. Phys.}\ }\textbf {\bibinfo {volume} {48}},\ \bibinfo {pages} {174004} (\bibinfo {year} {2015})}\BibitemShut {NoStop}%
\bibitem [{\citenamefont {Christopher}\ \emph {et~al.}(2005)\citenamefont {Christopher}, \citenamefont {Shapiro},\ and\ \citenamefont {Brumer}}]{christopher_overlapping_2005}%
  \BibitemOpen
  \bibfield  {author} {\bibinfo {author} {\bibfnamefont {P.~S.}\ \bibnamefont {Christopher}}, \bibinfo {author} {\bibfnamefont {M.}~\bibnamefont {Shapiro}},\ and\ \bibinfo {author} {\bibfnamefont {P.}~\bibnamefont {Brumer}},\ }\href {https://doi.org/10.1063/1.2000260} {\bibfield  {journal} {\bibinfo  {journal} {The Journal of Chemical Physics}\ }\textbf {\bibinfo {volume} {123}},\ \bibinfo {pages} {064313} (\bibinfo {year} {2005})}\BibitemShut {NoStop}%
\bibitem [{\citenamefont {Christopher}\ \emph {et~al.}(2006{\natexlab{a}})\citenamefont {Christopher}, \citenamefont {Shapiro},\ and\ \citenamefont {Brumer}}]{christopher_quantum_2006}%
  \BibitemOpen
  \bibfield  {author} {\bibinfo {author} {\bibfnamefont {P.~S.}\ \bibnamefont {Christopher}}, \bibinfo {author} {\bibfnamefont {M.}~\bibnamefont {Shapiro}},\ and\ \bibinfo {author} {\bibfnamefont {P.}~\bibnamefont {Brumer}},\ }\href {https://doi.org/10.1063/1.2346684} {\bibfield  {journal} {\bibinfo  {journal} {The Journal of Chemical Physics}\ }\textbf {\bibinfo {volume} {125}},\ \bibinfo {pages} {124310} (\bibinfo {year} {2006}{\natexlab{a}})}\BibitemShut {NoStop}%
\bibitem [{\citenamefont {Christopher}\ \emph {et~al.}(2006{\natexlab{b}})\citenamefont {Christopher}, \citenamefont {Shapiro},\ and\ \citenamefont {Brumer}}]{christopher_efficient_2006}%
  \BibitemOpen
  \bibfield  {author} {\bibinfo {author} {\bibfnamefont {P.~S.}\ \bibnamefont {Christopher}}, \bibinfo {author} {\bibfnamefont {M.}~\bibnamefont {Shapiro}},\ and\ \bibinfo {author} {\bibfnamefont {P.}~\bibnamefont {Brumer}},\ }\href {https://doi.org/10.1063/1.2196888} {\bibfield  {journal} {\bibinfo  {journal} {The Journal of Chemical Physics}\ }\textbf {\bibinfo {volume} {124}},\ \bibinfo {pages} {184107} (\bibinfo {year} {2006}{\natexlab{b}})}\BibitemShut {NoStop}%
\bibitem [{\citenamefont {Tscherbul}\ and\ \citenamefont {Brumer}(2014{\natexlab{b}})}]{tscherbul_long-lived_2014}%
  \BibitemOpen
  \bibfield  {author} {\bibinfo {author} {\bibfnamefont {T.~V.}\ \bibnamefont {Tscherbul}}\ and\ \bibinfo {author} {\bibfnamefont {P.}~\bibnamefont {Brumer}},\ }\href {https://doi.org/10.1103/PhysRevLett.113.113601} {\bibfield  {journal} {\bibinfo  {journal} {Phys. Rev. Lett.}\ }\textbf {\bibinfo {volume} {113}},\ \bibinfo {pages} {113601} (\bibinfo {year} {2014}{\natexlab{b}})}\BibitemShut {NoStop}%
\bibitem [{\citenamefont {Dodin}\ \emph {et~al.}(2016{\natexlab{a}})\citenamefont {Dodin}, \citenamefont {Tscherbul},\ and\ \citenamefont {Brumer}}]{dodin_quantum_2016}%
  \BibitemOpen
  \bibfield  {author} {\bibinfo {author} {\bibfnamefont {A.}~\bibnamefont {Dodin}}, \bibinfo {author} {\bibfnamefont {T.~V.}\ \bibnamefont {Tscherbul}},\ and\ \bibinfo {author} {\bibfnamefont {P.}~\bibnamefont {Brumer}},\ }\href {https://doi.org/10.1063/1.4954243} {\bibfield  {journal} {\bibinfo  {journal} {J. Chem. Phys.}\ }\textbf {\bibinfo {volume} {144}},\ \bibinfo {pages} {244108} (\bibinfo {year} {2016}{\natexlab{a}})}\BibitemShut {NoStop}%
\bibitem [{\citenamefont {Augenbraun}\ \emph {et~al.}(2023)\citenamefont {Augenbraun}, \citenamefont {Anderegg}, \citenamefont {Hallas}, \citenamefont {Lasner}, \citenamefont {Vilas},\ and\ \citenamefont {Doyle}}]{augenbraun_direct_2023}%
  \BibitemOpen
  \bibfield  {author} {\bibinfo {author} {\bibfnamefont {B.~L.}\ \bibnamefont {Augenbraun}}, \bibinfo {author} {\bibfnamefont {L.}~\bibnamefont {Anderegg}}, \bibinfo {author} {\bibfnamefont {C.}~\bibnamefont {Hallas}}, \bibinfo {author} {\bibfnamefont {Z.~D.}\ \bibnamefont {Lasner}}, \bibinfo {author} {\bibfnamefont {N.~B.}\ \bibnamefont {Vilas}},\ and\ \bibinfo {author} {\bibfnamefont {J.~M.}\ \bibnamefont {Doyle}},\ }\href {https://doi.org/10.48550/arXiv.2302.10161} {\bibinfo {title} {Direct {Laser} {Cooling} of {Polyatomic} {Molecules}}} (\bibinfo {year} {2023}),\ \Eprint {https://arxiv.org/abs/2302.10161} {arXiv:2302.10161} \BibitemShut {NoStop}%
\bibitem [{\citenamefont {Gallagher}(1994)}]{gallagher_rydberg_1994}%
  \BibitemOpen
  \bibfield  {author} {\bibinfo {author} {\bibfnamefont {T.}~\bibnamefont {Gallagher}},\ }\href@noop {} {\emph {\bibinfo {title} {Rydberg {Atoms}}}}\ (\bibinfo  {publisher} {Cambridge University Press},\ \bibinfo {year} {1994})\BibitemShut {NoStop}%
\bibitem [{\citenamefont {Tscherbul}\ and\ \citenamefont {Brumer}(2014{\natexlab{c}})}]{tscherbul_coherent_2014}%
  \BibitemOpen
  \bibfield  {author} {\bibinfo {author} {\bibfnamefont {T.~V.}\ \bibnamefont {Tscherbul}}\ and\ \bibinfo {author} {\bibfnamefont {P.}~\bibnamefont {Brumer}},\ }\href {https://doi.org/10.1103/PhysRevA.89.013423} {\bibfield  {journal} {\bibinfo  {journal} {Phys. Rev. A}\ }\textbf {\bibinfo {volume} {89}},\ \bibinfo {pages} {013423} (\bibinfo {year} {2014}{\natexlab{c}})}\BibitemShut {NoStop}%
\bibitem [{\citenamefont {Vilas}\ \emph {et~al.}(2023{\natexlab{a}})\citenamefont {Vilas}, \citenamefont {Hallas}, \citenamefont {Anderegg}, \citenamefont {Robichaud}, \citenamefont {Zhang}, \citenamefont {Dawley}, \citenamefont {Cheng},\ and\ \citenamefont {Doyle}}]{vilas_blackbody_2023}%
  \BibitemOpen
  \bibfield  {author} {\bibinfo {author} {\bibfnamefont {N.~B.}\ \bibnamefont {Vilas}}, \bibinfo {author} {\bibfnamefont {C.}~\bibnamefont {Hallas}}, \bibinfo {author} {\bibfnamefont {L.}~\bibnamefont {Anderegg}}, \bibinfo {author} {\bibfnamefont {P.}~\bibnamefont {Robichaud}}, \bibinfo {author} {\bibfnamefont {C.}~\bibnamefont {Zhang}}, \bibinfo {author} {\bibfnamefont {S.}~\bibnamefont {Dawley}}, \bibinfo {author} {\bibfnamefont {L.}~\bibnamefont {Cheng}},\ and\ \bibinfo {author} {\bibfnamefont {J.~M.}\ \bibnamefont {Doyle}},\ }\href {https://doi.org/10.1103/PhysRevA.107.062802} {\bibfield  {journal} {\bibinfo  {journal} {Phys. Rev. A}\ }\textbf {\bibinfo {volume} {107}},\ \bibinfo {pages} {062802} (\bibinfo {year} {2023}{\natexlab{a}})}\BibitemShut {NoStop}%
\bibitem [{\citenamefont {Vilas}\ \emph {et~al.}(2023{\natexlab{b}})\citenamefont {Vilas}, \citenamefont {Hallas}, \citenamefont {Anderegg}, \citenamefont {Robichaud}, \citenamefont {Zhang}, \citenamefont {Dawley}, \citenamefont {Cheng},\ and\ \citenamefont {Doyle}}]{vilas_blackbody_2023-1}%
  \BibitemOpen
  \bibfield  {author} {\bibinfo {author} {\bibfnamefont {N.~B.}\ \bibnamefont {Vilas}}, \bibinfo {author} {\bibfnamefont {C.}~\bibnamefont {Hallas}}, \bibinfo {author} {\bibfnamefont {L.}~\bibnamefont {Anderegg}}, \bibinfo {author} {\bibfnamefont {P.}~\bibnamefont {Robichaud}}, \bibinfo {author} {\bibfnamefont {C.}~\bibnamefont {Zhang}}, \bibinfo {author} {\bibfnamefont {S.}~\bibnamefont {Dawley}}, \bibinfo {author} {\bibfnamefont {L.}~\bibnamefont {Cheng}},\ and\ \bibinfo {author} {\bibfnamefont {J.~M.}\ \bibnamefont {Doyle}},\ }\href {https://doi.org/10.1103/PhysRevA.107.062802} {\bibfield  {journal} {\bibinfo  {journal} {Phys. Rev. A}\ }\textbf {\bibinfo {volume} {107}},\ \bibinfo {pages} {062802} (\bibinfo {year} {2023}{\natexlab{b}})}\BibitemShut {NoStop}%
\bibitem [{\citenamefont {Axelrod}\ and\ \citenamefont {Brumer}(2018)}]{axelrod_efficient_2018}%
  \BibitemOpen
  \bibfield  {author} {\bibinfo {author} {\bibfnamefont {S.}~\bibnamefont {Axelrod}}\ and\ \bibinfo {author} {\bibfnamefont {P.}~\bibnamefont {Brumer}},\ }\href {https://doi.org/10.1063/1.5041005} {\bibfield  {journal} {\bibinfo  {journal} {J. Chem. Phys.}\ }\textbf {\bibinfo {volume} {149}},\ \bibinfo {pages} {114104} (\bibinfo {year} {2018})}\BibitemShut {NoStop}%
\bibitem [{\citenamefont {Axelrod}\ and\ \citenamefont {Brumer}(2019)}]{axelrod_multiple_2019}%
  \BibitemOpen
  \bibfield  {author} {\bibinfo {author} {\bibfnamefont {S.}~\bibnamefont {Axelrod}}\ and\ \bibinfo {author} {\bibfnamefont {P.}~\bibnamefont {Brumer}},\ }\href {https://doi.org/10.1063/1.5099969@jcp.2019.OSQD2019.issue-1} {\bibfield  {journal} {\bibinfo  {journal} {J. Chem. Phys.}\ }\textbf {\bibinfo {volume} {151}},\ \bibinfo {pages} {014104} (\bibinfo {year} {2019})}\BibitemShut {NoStop}%
\bibitem [{\citenamefont {Dodin}\ and\ \citenamefont {Brumer}(2021)}]{dodin_generalized_2021}%
  \BibitemOpen
  \bibfield  {author} {\bibinfo {author} {\bibfnamefont {A.}~\bibnamefont {Dodin}}\ and\ \bibinfo {author} {\bibfnamefont {P.}~\bibnamefont {Brumer}},\ }\href {https://doi.org/10.1103/PRXQuantum.2.030302} {\bibfield  {journal} {\bibinfo  {journal} {PRX Quantum}\ }\textbf {\bibinfo {volume} {2}},\ \bibinfo {pages} {030302} (\bibinfo {year} {2021})}\BibitemShut {NoStop}%
\bibitem [{\citenamefont {Trushechkin}(2021)}]{trushechkin_unified_2021}%
  \BibitemOpen
  \bibfield  {author} {\bibinfo {author} {\bibfnamefont {A.}~\bibnamefont {Trushechkin}},\ }\href {https://doi.org/10.1103/PhysRevA.103.062226} {\bibfield  {journal} {\bibinfo  {journal} {Phys. Rev. A}\ }\textbf {\bibinfo {volume} {103}},\ \bibinfo {pages} {062226} (\bibinfo {year} {2021})}\BibitemShut {NoStop}%
\bibitem [{\citenamefont {Scully}\ and\ \citenamefont {Zubairy}(1997)}]{scully_quantum_1997}%
  \BibitemOpen
  \bibfield  {author} {\bibinfo {author} {\bibfnamefont {M.~O.}\ \bibnamefont {Scully}}\ and\ \bibinfo {author} {\bibfnamefont {M.~S.}\ \bibnamefont {Zubairy}},\ }\href@noop {} {\emph {\bibinfo {title} {Quantum {Optics}}}}\ (\bibinfo  {publisher} {Cambridge University Press},\ \bibinfo {year} {1997})\BibitemShut {NoStop}%
\bibitem [{\citenamefont {Dodin}\ \emph {et~al.}(2016{\natexlab{b}})\citenamefont {Dodin}, \citenamefont {Tscherbul},\ and\ \citenamefont {Brumer}}]{dodin_coherent_2016}%
  \BibitemOpen
  \bibfield  {author} {\bibinfo {author} {\bibfnamefont {A.}~\bibnamefont {Dodin}}, \bibinfo {author} {\bibfnamefont {T.~V.}\ \bibnamefont {Tscherbul}},\ and\ \bibinfo {author} {\bibfnamefont {P.}~\bibnamefont {Brumer}},\ }\href {https://doi.org/10.1063/1.4972140} {\bibfield  {journal} {\bibinfo  {journal} {The Journal of Chemical Physics}\ }\textbf {\bibinfo {volume} {145}},\ \bibinfo {pages} {244313} (\bibinfo {year} {2016}{\natexlab{b}})}\BibitemShut {NoStop}%
\bibitem [{\citenamefont {Koyu}\ and\ \citenamefont {Tscherbul}(2018)}]{koyu_long-lived_2018}%
  \BibitemOpen
  \bibfield  {author} {\bibinfo {author} {\bibfnamefont {S.}~\bibnamefont {Koyu}}\ and\ \bibinfo {author} {\bibfnamefont {T.~V.}\ \bibnamefont {Tscherbul}},\ }\href {https://doi.org/10.1103/PhysRevA.98.023811} {\bibfield  {journal} {\bibinfo  {journal} {Phys. Rev. A}\ }\textbf {\bibinfo {volume} {98}},\ \bibinfo {pages} {023811} (\bibinfo {year} {2018})}\BibitemShut {NoStop}%
\bibitem [{Note1()}]{Note1}%
  \BibitemOpen
  \bibinfo {note} {This condition guarantees that the $\phi $ autocorrelation function decays to a constant value and thus does not induce residual dynamics.}\BibitemShut {Stop}%
\end{thebibliography}%
\end{document}